\documentclass[aps, prc, 12pt, nofootinbib, showpacs, superscriptaddress, tightenlines, groupedaddress]{revtex4-1}
\usepackage{amsmath,amssymb,amsbsy,bm}
\usepackage{graphicx}
\usepackage{comment}
\usepackage{float}
\usepackage[colorlinks=true,linkcolor=blue,citecolor=blue,urlcolor=blue]{hyperref}
\usepackage[margin=0.75in]{geometry}

\usepackage[table]{xcolor}
\definecolor{purple}{rgb}{0.52, 0., 0.52}

\begin{document}

\title{Charge balance functions for heavy-ion collisions at energies
   available at the CERN Large Hadron Collider}
\author{Scott Pratt}
\affiliation{Department of Physics and Astronomy and National Superconducting Cyclotron Laboratory\\
Michigan State University, East Lansing, MI 48824~~USA}
\author{Christopher Plumberg}
\affiliation{Illinois Center for Advanced Studies of the Universe, Department of Physics, University of Illinois at Urbana-Champaign, Urbana, IL 61801, USA}
\date{\today}


\begin{abstract}
Heavy-ion collisions at the LHC provide the conditions to investigate regions of quark-gluon plasma that reach higher temperatures and that persist for longer periods of time compared to collisions at the Relativistic Heavy Ion Collider. This extended duration allows correlations from charge conservation to better separate during the quark-gluon plasma phase, and thus be better distinguished from correlations that develop during the hadron phase or during hadronization. In this study charge balance functions binned by relative rapidity and azimuthal angle and indexed by species are considered. A detailed theoretical model that evolves charge correlations throughout the entirety of an event is compared to preliminary results from the ALICE Collaboration. The comparison with experiment provides insight into the evolution of the chemistry and diffusivity during the collision. A ratio of balance functions is proposed to better isolate the effects of diffusion and thus better constrain the diffusivity.
\end{abstract}

\maketitle

\section{Introduction}

Charge is locally conserved in heavy-ion collisions. This includes electric charge $Q$, baryon number $B$, and strangeness $S$, which may be expressed equivalently in terms of the net number of up, down and strange quarks. Once balancing charge pairs are created, they separate diffusively. For charge pairs created in the initial thermalization of the fireball, the characteristic separations might exceed a unit of rapidity along the beam direction, and transversely, the charges might even find themselves on opposite sides of the fireball. When coupled with the explosive collective flow present in heavy-ion collisions, this spatial separation leads to the balancing charges being thrust in different directions, thereby creating momentum-space correlations between charge pairs. 

These correlations can be studied by measuring \textit{charge balance functions}, which represent the probability, given the observation of a charge $q$, of seeing its balancing charge $-q$ at some relative rapidity $\Delta y$, relative azimuthal angle $\Delta\phi$, or relative momentum $Q_{\rm inv}$. For example \cite{Bass:2000az},
\begin{eqnarray}\label{eq:balancedef}
B(\Delta y,\Delta \phi)&=&\int d\phi_1 dy_1 d\phi_2 dy_2~\delta(y_1-y_2-\Delta y) \delta(\phi_1-\phi_2-\Delta\phi)\\
\nonumber
&&\times\left\{
 \frac{1}{2P_+}\left[P_{-+}(\phi_1,y_1;\phi_2,y_2)-P_{++}(\phi_1,y_1;\phi_2,y_2)\right] \right.\nonumber\\
\nonumber
&&\left. +\frac{1}{2P_-}\left[P_{+-}(\phi_1,y_1;\phi_2,y_2)-P_{--}(\phi_1,y_1;\phi_2,y_2)\right]\right\},
\end{eqnarray}
where $P_\pm=P_\pm(\phi_1,y_1)$ and, for example, $P_{+-}(\phi_1,y_1;\phi_2,y_2)/P_+$ represents the conditional probability density for finding a negative charge at $(\phi_2,y_2)$, given finding a positive charge at $(\phi_1,y_1)$.

Balance functions can also be indexed by species. For example, $B_{K|p}$ would require that the particle at $\vec{p}_2$ is a proton or antiproton and that the particle at $\vec{p}_1$ is a proton or antiproton \cite{Pratt:2012dz},
\begin{eqnarray}\label{eq:genbalancedef}
B_{K|p}(\Delta y,\Delta \phi)&=&\int d\phi_1 dy_1 d\phi_2 dy_2~\delta(y_1-y_2-\Delta y) \delta(\phi_1-\phi_2-\Delta\phi)\\
\nonumber
&&
\left\{
 \frac{1}{P_{p}}      \left[P_{K^-,p}-P_{K^+,p}\right]
+\frac{1}{2P_{\bar{p}}}\left[P_{K^+,\bar{p}}-P_{K^-,\bar{p}}\right]\right\},
\end{eqnarray}
and the arguments of the $P_{K^-,p}(\phi_1,y_1;\phi_2,y_2)$ etc. are suppressed for readability.  $B_{K|p}$ is thus the conditional probability of observing a charge on a kaon or antikaon, given that the opposite charge was observed on a $p$ or $\bar{p}$.

Charge balance functions have been measured for a variety of binnings, indexed by a variety of species, and have yielded valuable insights into the different stages of charge production in nuclear collisions \cite{Wang:2012jua,Abelev:2010ab,Li:2011zzx,Adams:2003kg,Aggarwal:2010ya,Alt:2004gx,Alt:2007hk,Adamczyk:2015yga,Wang:2011za,STAR:2011ab,Westfall:2004cq,Pan:2018dsq,JinjinPanThesis,Alam:2017iom,Weber:2013fla,Abelev:2013csa,Weber:2012ut}. For instance, because most electric charge is created during hadronization or in the subsequent decays, most balancing charges are emitted with similar momenta, with the increasingly strong collective flow for more central collisions focusing the charges into increasingly similar relative rapidities and relative angles \cite{Adams:2003kg}. Because pions carry only electric charge, and because most of the electric charge is created late in the collision, $B_{\pi|\pi}(\Delta y)$ thus tends to become narrower for more central collisions. In contrast, strangeness is mostly created in the earliest stages of the collision, as the quark-gluon plasma (QGP) is first formed, so that $B_{K|K}$ does not narrow with increasing centrality.  Constructing charge balance functions for a variety of species and/or charges thus provides sensitivity to different stages of the system's evolution.  Avoiding those species which require reconstruction from weak decays or measurement of neutrons, one is left with pions, kaons, and protons for constraining the correlations of $B$, $Q$, and $S$.  This yields a minimally complete set of six balance functions, $B_{\pi|\pi}$, $B_{\pi|K}$, $B_{\pi|p}$, $B_{K|K}$, $B_{K|p}$ and $B_{p|p}$, each of which can be binned by relative rapidity, azimuthal angle or relative momentum.\footnote{In principle, one may also consider the formally distinct balance functions, $B_{K|\pi}$, $B_{p|\pi}$, and $B_{p|K}$ as well.  However, the six balance functions already mentioned provide sufficient discriminating power for the results we will present here.}

For the purpose of modeling the charge balance functions measured in nuclear collisions, detailed calculations are required in order to evolve the charge-charge correlation functions throughout the QGP phase and then project these correlations onto the species-dependent balance functions. As reviewed in the next section, the charge-charge correlations, $\langle\rho_a(x)\rho_b(x')\rangle$, can be modeled by solving the diffusion equation with a source function given by the evolution of the charge susceptibility, $\chi_{ab}(x)$. State-of-the-art models of high-energy heavy ion collisions employ hydrodynamic models to describe the QGP phase, and switch to hadronic simulations for the hadron phase. At the hyper-surface separating these two descriptions, the charge-charge correlation function must be translated into correlations indexed by hadron species based on statistical arguments.  The accurate modeling of conserved charge dynamics thus raises a number technical challenges for constructing and propagating charge correlations throughout the collision lifetime which have been addressed previously in the context of heavy-ion collisions at RHIC \cite{Pratt:2017oyf, Pratt:2018ebf,Pratt:2019pnd}.

Experimentally, measurements at RHIC have quantitatively confirmed the resulting model-based expectations that $B_{\pi|\pi}(\Delta y)$ narrows with centrality while $B_{K|K}$ and $B_{p|p}$ do not, and that for central collisions charge balance functions for kaons and protons are broader than those for pions \cite{Aggarwal:2010ya,Pratt:2015jsa,Pratt:2018ebf}.  By assuming that the matter was chemically equilibrated at an early time, and that the diffusivity matched that of lattice calculations, the model was able to explain the width in relative rapidity of all the balance functions measured by STAR \cite{Pratt:2019pnd}. Charge balance functions therefore provide some of the most powerful and direct evidence available for the creation of a state of matter with the requisite number of light quarks to attain chemical equilibrium. In addition, binning by azimuthal angle has been shown to provide the means to constrain the diffusivity \cite{Pratt:2019pnd}: by focusing on transverse separation in $KK$ and $pp$ balance functions, one obtains superior insight and largely avoids ambiguities related to how far charges have already separated during the first 1 fm/$c$, and because these species are especially sensitive to charge created early.  Furthermore, although species-indexed charge balance functions binned by azimuthal angle have not yet been measured at RHIC, those analyses might appear soon.  

The principal goal of this study is to extend the ideas above to the modeling, analysis and interpretation of data from the LHC. Collisions at the LHC attain approximately double the initial energy density as those at RHIC, and more time is spent in the QGP phase. One thus expects increased leverage and constraining power, both for investigating the chemical evolution of the matter created in the collision and for extracting the diffusivity of the QGP. The same model applied to RHIC collisions in \cite{Pratt:2018ebf,Pratt:2017oyf,Pratt:2019pnd} will be applied here. Preliminary results from the ALICE Collaboration have appeared in the last year and will be compared to model results here \cite{Pan:2018dsq,JinjinPanThesis}.

The remainder of this paper is organized as follows.  The theory of charge correlations and their evolution through the hydrodynamic stage is reviewed in Sec. \ref{sec:hydro}. Section \ref{sec:cascade} shows how those correlations are projected onto and evolved through the hadronic simulation, and Sec. \ref{sec:model} reviews the algorithms and methods of the calculation. Section \ref{sec:sigma} investigates sensitivity to the initial conditions at the time which the hydrodynamic description is instantiated. The comparison to preliminary results from ALICE makes a case that chemical equilibrium was approached early in the collision. The analysis also has implications of how thermalization might have been reached at early times. Section \ref{sec:diffusion} emphasizes the extraction of diffusivity of the QGP. By comparing to preliminary ALICE results it appears that the diffusivity is in the neighborhood of that from lattice calculations, but some caveats remain. Also, in Sec. \ref{sec:diffusion} a novel observable is proposed from which one can more robustly extract the diffusivity. By taking the ratio of angular harmonics of the balance function and by confining the observable outside the small relative rapidity region, one can avoid some of the caveats from interpreting measurements of $B(\Delta\phi)$ alone. A summary, along with a discussion of prospects for future analyses, is presented in Sec. \ref{sec:summary}.


\section{Theory: Charge correlations in coordinate space}\label{sec:hydro}

In order to make the paper more self contained this section reviews the formalism and methods discussed in detail in \cite{Pratt:2017oyf}.

\subsection{General formalism}
Calculating charge balance functions requires first computing the evolution of the charge-charge correlation,
\begin{eqnarray}
C_{ab}(\vec{r}_1,\vec{r}_2,\tau)&=&\langle\rho_a(\vec{r}_1,\tau)\rho_b(\vec{r}_2,\tau)\rangle.
\end{eqnarray}
Here, the charge density of charge type $a$ (where $a$ is up, down or strange) is represented by $\rho_a$. If one were considering a system where the average charge were non-zero, $\rho_a$ would be replaced by $\delta\rho_a=\rho_a-\langle\rho_a\rangle$, and the following discussion would be accordingly modified. For a non-interacting quark gas, ignoring Fermi statistics, there are no correlations between different quarks and the correlation becomes
\begin{eqnarray}
C_{ab}(\vec{r}_1,\vec{r}_2)&=&[n_a(\vec{r}_1)+n_{\bar{a}}(\vec{r}_1)]\delta(\vec{r}_1-\vec{r}_2)\delta_{ab}.
\end{eqnarray}
Here $n_a$ and $n_{\bar a}$ are the densities of the quark and antiquarks of flavor $a$. For a gas this correlation is non-zero only when $\rho_a$ and $\rho_b$ refer to the same charge, and thus represents the correlation of a charge with itself. This leads to Poissonian multiplicity fluctuations. Including Fermi statistics leads to a negative correlation at relative distances $r\lesssim 1/k_f$, but that will be neglected for the moment. For a non-interacting hadron gas, correlations occur only for charges within the same hadron, so if the equilibrium density of each hadron species $h$ is $n_h$, then
\begin{eqnarray}\label{eq:chihadgas}
C_{ab}(\vec{r}_1,\vec{r}_2)&=&\sum_h n_h(\vec{r}_1)q_{ha}q_{hb}\delta(\vec{r}_1-\vec{r}_2),
\end{eqnarray}
where $q_{ha}$ is the charge of flavor $a$ on hadron species $h$. Because hadrons might be composed of multiple quarks, off-diagonal elements are  non-zero. For example, the contribution from pions to $C_{ud}$ is $-(n_{\pi^+}+n_{\pi^-})\delta(\vec{r}_1-\vec{r}_2)$. More realistically, the delta function would be replaced by a function whose strength extends over the size of a hadron, but would still integrate to unity as a delta function. So the delta function here can be considered as some short-range function that integrates to unity. Any short-range clustering should be encompassed by the ``delta'' function.

The charge susceptibility, or charge fluctuation, for a large volume $V$ is given by the charge correlation,
\begin{eqnarray}
\chi_{ab}&=&\left\langle\frac{Q_aQ_b}{V}\right\rangle\\
\nonumber
&=&\frac{1}{V}\int d^3rd^3r_1~C_{ab}(\vec{r}_1-\vec{r}_2).
\end{eqnarray}
If the correlation is purely local, then it is proportional to a delta function:
\begin{eqnarray}
C_{ab}(\vec{r}_1-\vec{r}_2)&=&\chi_{ab}(\vec{r}_1)\delta(\vec{r}_1-\vec{r}_2).
\end{eqnarray}

The preceding expressions have all been written for equilibrated gases in infinite systems. The balancing charges were neglected, which is justified if the system is large and the balancing charges have had sufficient time to diffuse over an arbitrarily large volume. If the net charge of the system is fixed within a finite volume, there must exist an additional correlation enforcing that fact that for any charge at $\vec{r}_1$ there exists another, opposite charge somewhere within a distance scale defined by how far the balancing charge might have diffused since the charge was initially created. This additional correlation, $C'_{ab}(\vec{r}_1-\vec{r}_2)$, must integrate to $-\chi_{ab}$ if the net charge is fixed,
\begin{eqnarray}\label{eq:nonlocaldef}
C_{ab}(\vec{r}_1,\vec{r}_2)&=&\chi_{ab}(\vec{r}_1)\delta(\vec{r}_1-\vec{r}_2)+C'_{ab}(\vec{r}_1,\vec{r}_2),\\
\nonumber
\int d^3r_2~C'_{ab}(\vec{r}_1,\vec{r}_2)&=&-\chi_{ab}(\vec{r}_1).
\end{eqnarray}
Here, $\chi_{ab}$ represents the strength of the local correlation and need not be the equilibrated susceptibility unless one is assuming chemical equilibrium. The balancing correlation, $C'_{ab}$, should spread diffusively over time,
\begin{eqnarray}
\partial_t C'_{ab}(\vec{r}_1,\vec{r}_2,t)&=&
-(\nabla_1\cdot\vec{v}_1+\vec{v}_1\cdot\nabla_1+\nabla_2\cdot\vec{v}_2+\vec{v}_2\cdot\nabla_2)C'_{ab}(\vec{r}_1,\vec{r}_2,t)\\
\nonumber
&&+D(\vec{r}_1,t)\nabla_1^2C'_{ab}(\vec{r}_1,\vec{r}_2,t)+D(\vec{r}_1,t)\nabla_1^2C'_{ab}(\vec{r}_1,\vec{r}_2,t)+S_{ab}(\vec{r}_1,t)\delta(\vec{r}_1-\vec{r}_2).
\end{eqnarray}
Here, $\vec{v}_1$ and $\vec{v}_2$ are the collective flow velocities of the matter at $\vec{r}_1$ and $\vec{r}_2$. The source function is determined from the above constraint of the integral of $C'_{ab}$ in Eq. (\ref{eq:nonlocaldef}) \cite{Pratt:2017oyf,Pratt:2018ebf},
\begin{eqnarray}\label{eq:Sabofchi}
S_{ab}(\bm{r},t)&=&(\partial_t+\bm{v}\cdot\nabla+\nabla\cdot\bm{v})\chi_{ab}(\bm{r},t).
\end{eqnarray}
If entropy is conserved, this can be restated as
\begin{eqnarray}
S_{ab}(\vec{r},t)&=&-s(\partial_t+\bm{v}\cdot\nabla)\frac{\chi_{ab}(\vec{r},t)}{s(\vec{r},t)}.
\end{eqnarray}
For a gas of massless quarks that expands isentropically, the net number of quarks does not change within a volume of fixed entropy. The ratio $\chi_{ab}/s$ then stays constant and the source function is zero.

\subsection{Charge correlations in relativistic heavy-ion collisions}
For relativistic heavy-ion collisions, a good fraction of the total charge is created in the early instants when the system transforms from a vacuum to a QGP. If the matter is established at some time $\tau_0$ with a local susceptibility $\chi^{(0)}_{ab}$, the initial correlation $C'_{ab}$ must integrate to $-\chi^{(0)}_{ab}$. Given that the time scale and mechanism for initial charge creation is not well understood, one must choose some form for the initial correlation. One expects $C'_{ab}(\tau_0)$ to be highly correlated in transverse coordinate space, but one can only speculate about the spread along the longitudinal (beam) direction. Even a small separation, e.g. 0.5 fm, is substantial if it is established at short times due to the large initial velocity gradients along the beam axis. The separation would then rapidly grow as the fluid elements containing the two particles flow apart, even if diffusion were neglected.

The hadronization region is also characterized by large changes of $\chi_{ab}/s$, and thus $S_{ab}$ becomes large in these regions. To a rough approximation, then, the correlation observed in the final stages of a relativistic heavy ion collisions is characterized by two sources. The first is the initial charge production during the first $\sim$1 fm/$c$ of the collision, when the QGP was created, and the second is the hadronization region. The first scale might subtend $\gtrsim 1$ unit of spatial rapidity, $\eta_s$, whereas the second source of correlation should be characterized by $\Delta\eta_s\lesssim 0.3$. 

\begin{figure}
\centerline{\includegraphics[width=0.48\textwidth]{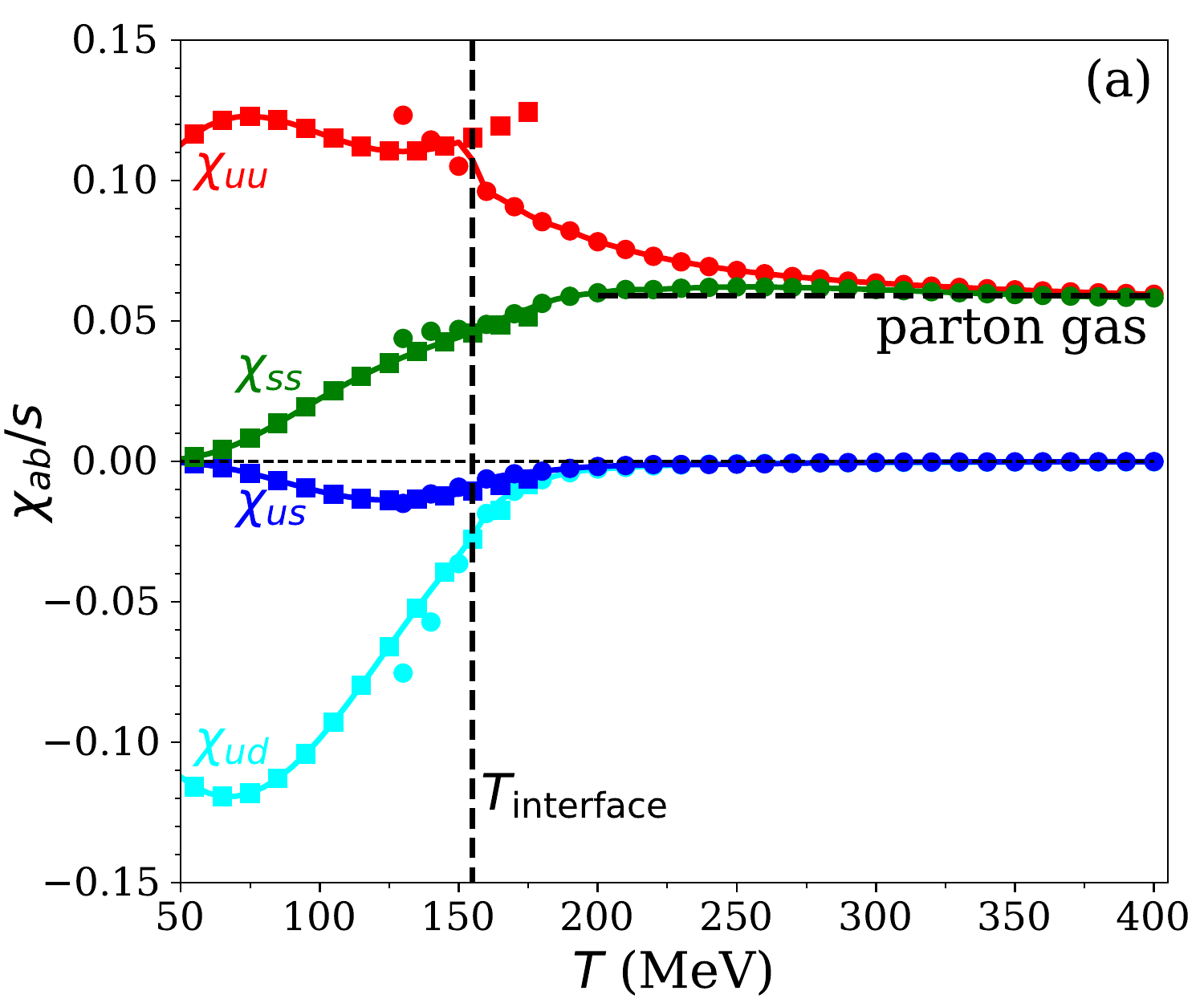}~~\includegraphics[width=0.48\textwidth]{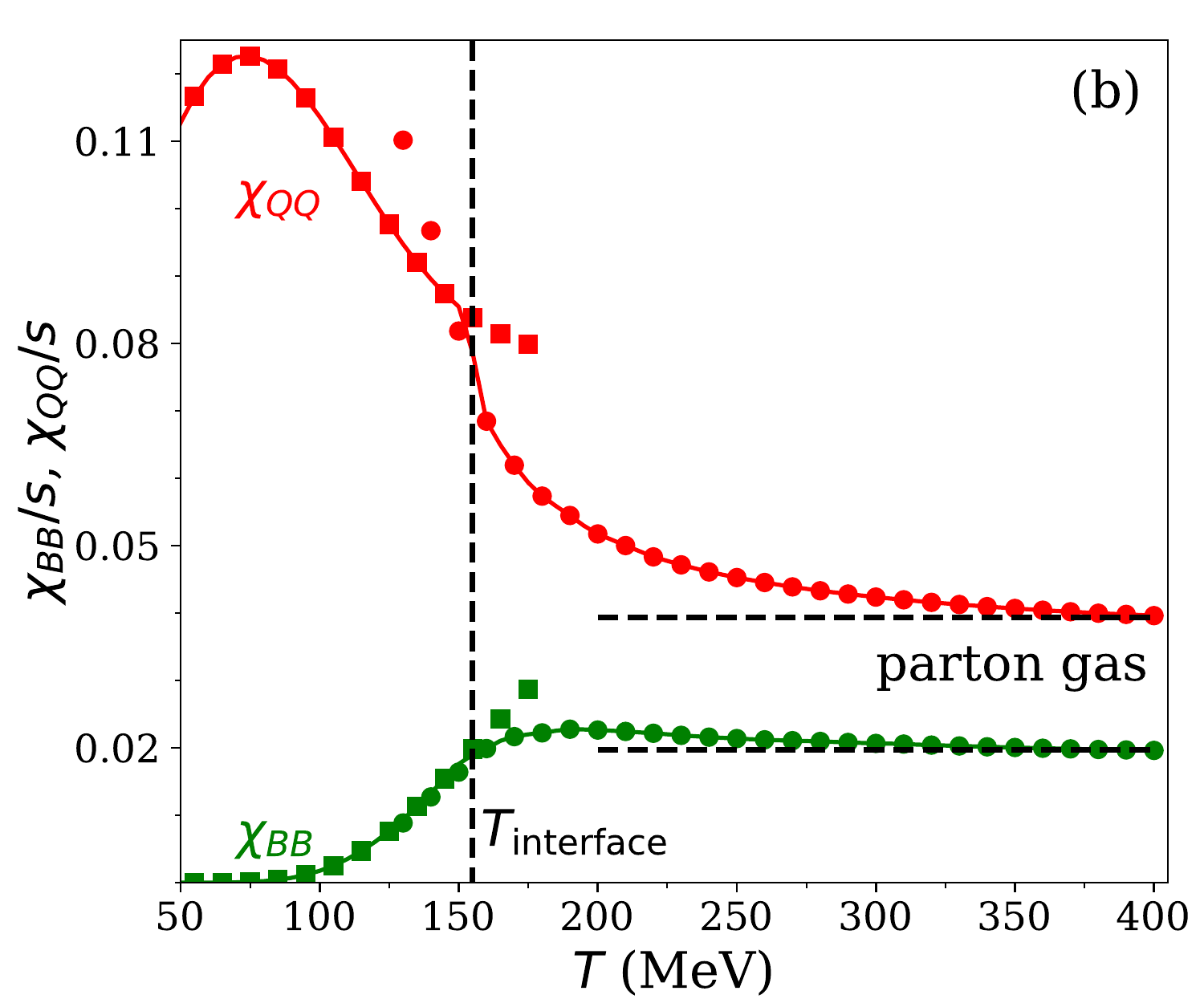}}
\caption{\label{fig:chi}(color online)
Panel (a): The charge susceptibility scaled by the entropy density indexed in terms of up, down and strange as calculated in lattice gauge theory (circles) and for a hadron gas (squares), with solid lines showing the interpolated values used in this paper. The susceptibility $\chi_{uu}=\chi_{dd}$ udergoes a sharp rise as the matter cools and traverses the hadronization region. This rise provides a strong surge to the source function for the correlation function $C'_{uu}$. In contrast, there is a small fall for $\chi_{ss}$ in the same region due to the relatively large mass penalty for strange quarks in a hadron phase. Non-diagonal terms disappear when hadrons dissolve into the QGP. The dashed line shows the temperature used here to transition from hydrodynamics to a microscopic hadronic simulation.\\
Panel (b): The charge susceptibility from the same lattice calculations indexed by baryon number and electric charge demonstrates strong growth as the matter hadronizes. This feeds the electric charge correlation function at the end of the reaction which results in a narrow balance function for pions. In contrast, because of the large relative mass for baryons, the balance function indexed by baryon number stays steady throughout hadronization. Thus, the lack of a late-time surge to the source function for the baryon-baryon correlation function results in a broader correlation function for balance functions of protons as compared to that of pions.}
\end{figure}

In reality, the source function is non-zero during the entire evolution. Figure \ref{fig:chi} displays the ratio $\chi_{ab}/s$ from lattice calculations as a function of temperature \cite{Borsanyi:2011sw}. As the temperature falls below $\approx 200$ MeV the susceptibilities change rapidly. This is especially true for $\chi_{uu}/s=\chi_{dd}/s$ and for $\chi_{ud}/s$. The rise in $\chi_{uu}/s$ and $\chi_{dd}/s$ as the temperature falls can be understood from entropy arguments \cite{Bass:2000az}. In order to conserve entropy during hadronization the number of quasi-particles should fall, but not precipitously. In a QGP each quasi-particle carries a bit less than four units of entropy, and in a hadron gas the entropy per particle is close to five units. Thus, one expects the number of constituents to fall by a few tens of percent. However, each hadron carries at least two quarks, and because the contribution to $\chi_{ab}$ for hadron $h$ is proportional to $q_{ha}q_{hb}$, some hadrons contribute strongly. For example, the $\Delta^{++}$ contributes nine times as much to $\chi_{uu}$ as a single up quark. Thus, hadronization is accompanied by rapid production of quark-antiquark pairs. In an idealized QGP, $\chi_{ab}$ has no off-diagonal terms. In contrast, as shown in Eq. (\ref{eq:chihadgas}), a hadron gas has strong off-diagonal elements because the quasi-particles carry multiple quarks of different flavors. In fact, as $T\rightarrow 0$ the hadron gas becomes a pion gas and $\chi_{du}=-\chi_{uu}$. According to the lattice calculations displayed in Fig. \ref{fig:chi} the off-diagonal elements have largely vanished for temperatures near 185 MeV, suggesting hadrons do not exist above this temperature. However, the same lattice calculations show that $\chi_{uu}$ does not approach that of a QGP until temperatures are in the neighborhood of 250 MeV. This suggests that the temperature dependence of $\chi_{uu}=\chi_{dd}$ in this temperature range might be due to the emergence of gluonic degrees of freedom over a broader range of temperatures. Thus, the creation of $u\bar{u}$ and $d\bar{d}$ pairs as matter traverses this broader temperature window might be largely driven by the disappearance of gluonic degrees of freedom.

\begin{figure}
\centerline{\includegraphics[width=0.48\textwidth]{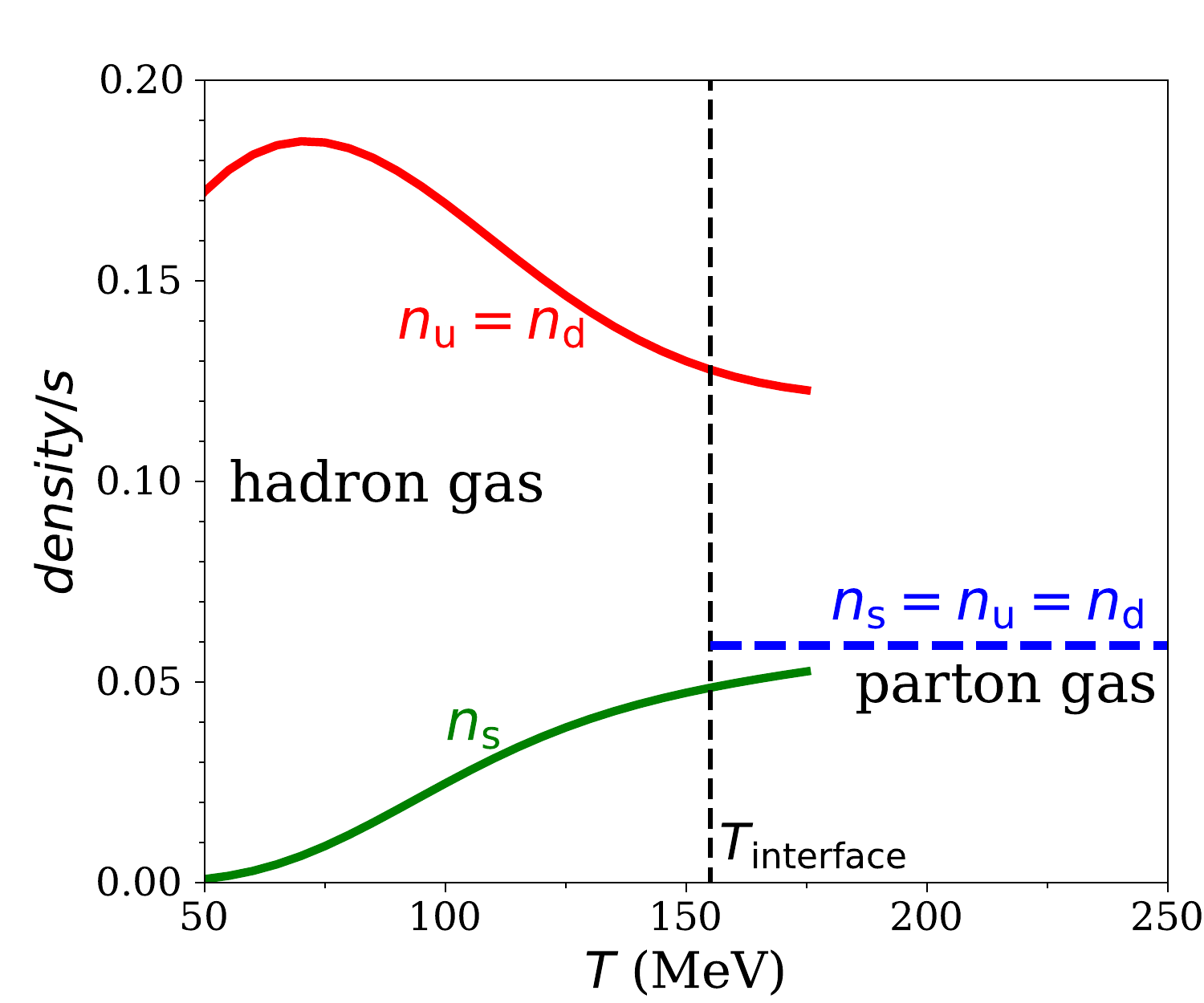}}
\caption{\label{fig:uds}(color online)
Using the nominal u,d,s content of hadrons, the quark density scaled by the entropy density of an equilibrated hadron gas is shown as a function of temperature. Also displayed are the same ratios for a massless parton gas. This corroborates the message from Fig. \ref{fig:chi}, that hadronization is accompanied by a large surge of $u,\bar{u},d$ and $\bar{d}$ quarks, while the number of $s$ and $\bar{s}$ quarks decreases modestly.}
\end{figure}

Figure \ref{fig:uds} provides an alternative insight into the chemical evolution of hadronization by showing the quark content of a hadron gas as a function of temperature. It displays the number density of up, down and strange quarks, referred to as $n_u$, $n_d$ and $n_s$. This differs from the charge as both $u$ and $\bar{u}$ quarks contribute positively to the quark density. Thus, a $\pi_0$ contributes to $n_u$ and $n_d$ even though the $\pi_0$ does not contribute to $\chi_{uu}$ or $\chi_{dd}$. By comparing to the ratio for the idealized QGP, one can see that as a QGP cools and hadronizes the number of up and down quarks nearly triples. In contrast, the number of strange quarks actually falls by $\approx 20$\%. This is expected given that the mass penalty for a strange quark is much lower for the QGP, $\approx 90$ MeV, as compared to the mass difference between a pion and a kaon, $\approx 350$ MeV. This difference provides critical leverage for validating the assumption that a chemically equilibrated QGP was created very early in the collision, and that the QGP persisted for a significant time. Experiments can measure the balance functions of kaons as a function of relative rapidity and compare it to that of pions. If the kaon balance function is largely influenced by the strange-quark correlation function it should be significantly broader than that of the pion correlation function. Indeed, even though pion balance functions are thermally broadened more so than their kaonic counterparts by thermal motion, STAR's measurements of kaon balance functions are significantly broader than those of pion balance functions.

Figures \ref{fig:chi} and \ref{fig:uds} display quantities indexed by the up, down and strange charge. From these, one can also generate correlations indexed by baryon number $B$ or electric charge $Q$. The baryon number and charge susceptibilities are
\begin{eqnarray}
\chi_{BB}&=&(2\chi_{uu}+2\chi_{ud}+\chi_{ss}+4\chi_{us})/9,\\
\nonumber
\chi_{QQ}&=&(5\chi_{uu}-4\chi_{ud}+\chi_{ss}-2\chi_{us})/9.
\end{eqnarray}
Figure \ref{fig:chi} also presents the baryon and charge susceptibilities in the same format as the susceptibilities indexed by $u,d,s$ in Fig. \ref{fig:chi}. The behavior of $\chi_{QQ}$ mimics the behavior of $\chi_{uu}=\chi_{dd}$. As the QGP cools and enters the hadronization region, $\chi_{BB}$ rises modestly. This is driven by the fact that baryons carry three quarks and contribute to $\chi_{BB}$ with nine times the strength as a single quark. However, as the temperature cools further, the large baryon mass suppresses $\chi_{BB}$ strongly as $T$ falls below 150 MeV. Thus, even though there is a strong increase in the number of up and down quarks during hadronization, $\chi_{BB}/s$ falls with decreasing temperature. To maintain chemical equilibrium, i.e. having the local correlation determined by the equilibrium susceptibility,  baryon annihilation is necessary. In realistic time scales, annihilation indeed occurs, but not at the pace needed to maintain chemical equilibrium.

When a susceptibility ratio falls the corresponding contribution to the source function is negative and results in a dip in the correlations function at small relative position. This then feeds a dip in the correlation function at small relative momentum. If this is due to annihilation, the denominator of the balance function is also reduced, and even though a dip has emerged in the correlation function, the integral of the balance function is unchanged, as it must be if charge is globally conserved. Thus, the width of the balance function spreads due to annihilation while the normalization is roughly unchanged. However, when including the effects of finite experimental acceptance the normalization will be reduced as more of the balance function's strength will be pushed outside the acceptance.

\section{Theory: Evolving correlations through the hydrodynamic interface and hadronic simulation}\label{sec:cascade}

Hydrodynamic modeling of heavy-ion collisions is limited to those parts of the evolution where the matter is locally thermalized, at least kinetically. Due to the large variation of masses in the hadronic phase, it is difficult to maintain kinetic equilibrium between the various species, let alone chemical equilibrium. Thus, in order to model the hadronic stage and the dissolution of the hydrodynamic medium, microscopic simulations are typically employed. The charge-charge correlations must then be transferred to the degrees of freedom of the simulation. This requires representing the quark charge correlations (indexed by $u,d,s$) in terms of hadron-hadron correlations (indexed by the species $h$). The method used in this study has been previously applied and explained in \cite{Pratt:2019pnd,Pratt:2018ebf}. After briefly reviewing this method, a discussion is presented here of how the integrated strength of the charge balance functions indexed by hadron species is constrained. These constraints play a critical role in understanding how discrepancies between experimental results from ALICE and model calculations presented later in this manuscript.

\subsection{Charge correlations and hadronization}
The charge correlation, $C'_{ab}(\vec{r}_1,\vec{r}_2)$, can be represented by pairs of tracer charges $\delta Q_1$ at $\vec{r}_1$ and $\delta Q_2$ at $\vec{r}_2$ in the differential volume elements $d^3r_1$ and $d^3r_2$, with
\begin{eqnarray}
C'_{ab}(\vec{r}_1,\vec{r}_2)d^3r_1d^3r_2&=&\delta Q_a\delta Q_b. 
\end{eqnarray}
When the volume hadronizes these differential charges translate into differential multiplicities of hadrons, $\delta N_h$. Here, we assume that $\delta N_h$ can be found according to statistical arguments. If a charge $\delta Q_a$ is spread statistically, it can be described by a small chemical potential, $\delta(\mu/T)$, where each species is affected through the chemical potential,
\begin{eqnarray}
\delta N_h&=&\langle N_h\rangle (e^{\delta\mu_aq_{ha}/T}-1)=\langle N_h\rangle \sum_a q_{ha}\delta(\mu_a/T).
\end{eqnarray}
Here, $q_{ha}$ is the charge of type $a$ on a hadrons species $h$ and $\langle N_h\rangle$ is the average number of the species within the volume. In turn, the charge $\delta Q_a$ must be generated by the charge carried by all the species,
\begin{eqnarray}
\delta Q_a&=&\sum_h q_{ha}\delta N_h\\
\nonumber
&=&\sum_{h,b}\langle N_h\rangle \sum_b q_{ha}q_{hb}\delta(\mu_b/T)\\
\nonumber
&=&\chi_{ab}V\delta(\mu_b/T).
\end{eqnarray}
Inverting this as a matrix equation for the chemical potential then gives the $\delta N_h$:
\begin{eqnarray}\label{eq:NhofQ}
\delta(\mu_a/T)&=&\frac{1}{V}\chi^{-1}_{ab}\delta Q_a,\\
\nonumber
\delta N_h&=&n_h\chi^{-1}_{ab}q_{ha}\delta Q_{b},
\end{eqnarray}
where $n_h$ is the equilibrium density of species $h$. Thus, the susceptibility, or charge fluctuation, provides the means to project the differential charge onto a differential hadron. For the calculations presented here the correlation in the hydrodynamic stage is represented by weighted pairs of differential charges in a Monte Carlo procedure. This correlation is then carried by weighted pairs of hadrons in the hadronic simulation. 

The corresponding balance functions indexed by hadron species are
\begin{eqnarray}
B_{h|h'}(p|p')&=&\frac{1}{2N_{h'}(p')}\langle[N_{\bar{h}}(p)-N_h(p)]N_{h'}(p')\rangle
+\frac{1}{2N_{\bar{h}'}(p')}\langle[N_h(p)-N_{\bar{h}}(p)]N_{\bar{h}'}(p')\rangle.
\label{Bhhp_definition}
\end{eqnarray}
Here $h$ and $\bar{h}$ refer to the positively or negatively charged hadron and its antiparticle. The labels $p$ and $p'$ could refer to momentum, rapidity, or azimuthal angle. Most commonly, $p'$ refers to any momentum that fits in the acceptance, and $h$ refers to some measure of relative momentum, e.g. the relative rapidity $y-y'$. This reduces the charge balance function to be some measure of relative momentum such as relative rapidity or relative angle. But, in principle, the balance function is a six-dimensional measurement if $p'$ is confined to a specific range of rapidity $y$, transverse momentum $p_t$ and azimuthal angle $\phi$.

\subsection{Partial normalization and resonance decays}
\begin{figure}
\centerline{\includegraphics[width=0.5\textwidth]{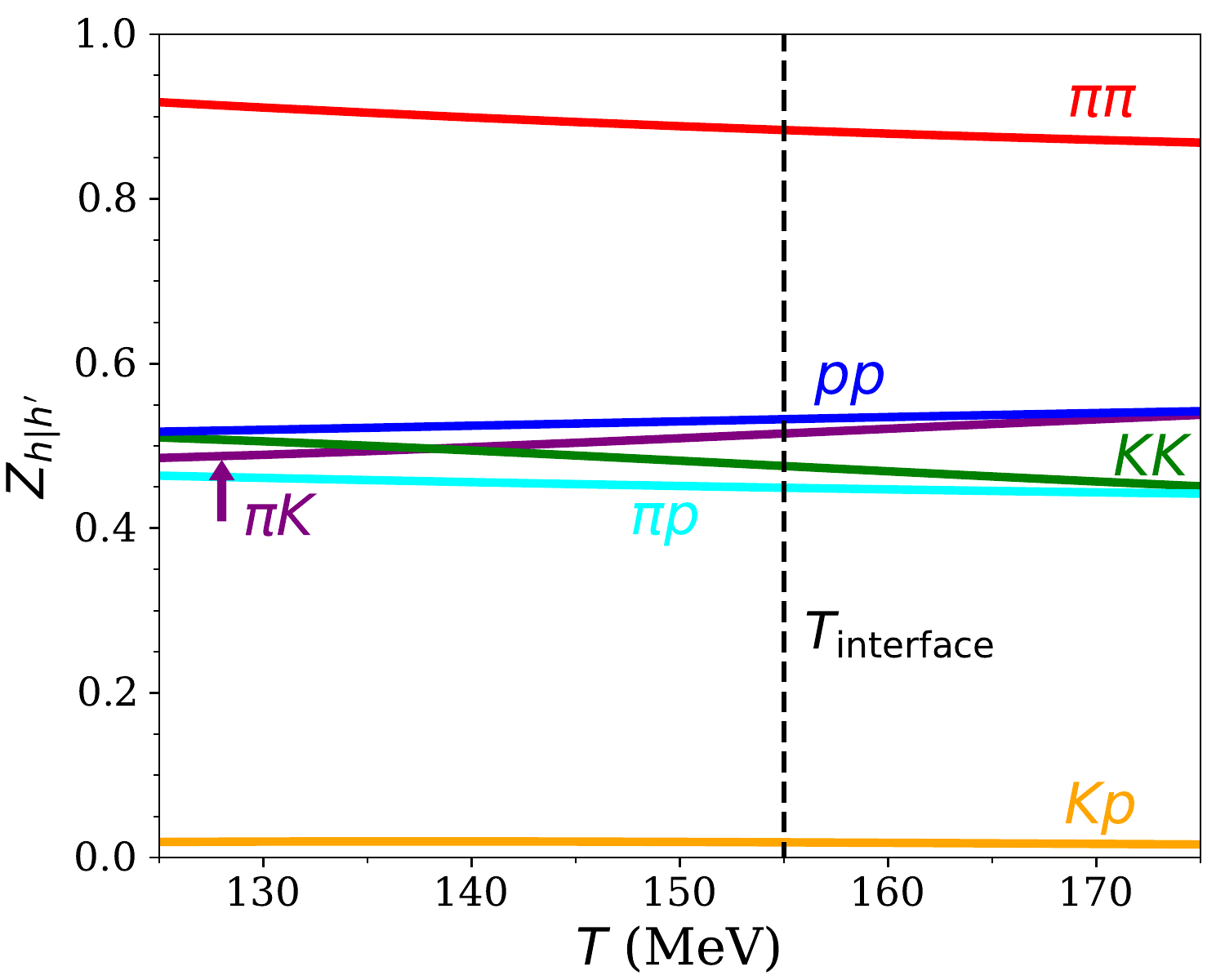}}
\caption{\label{fig:bfnorm}(color online)
The partial normalization of balance functions indexed by hadron species. $Z(h|h')$ gives the additional amount of opposite- vs. same-sign charged hadrons given the observation of a charged hadron. The indices $h$ and $h'$ require that the first measurement of is confined to mesons of species $h'$ or $\bar{h}'$ and that the second particle is either an $h$ or an $\bar{h}$. The normalization shows that for every charged pion, one can find nearly 90\% of the balancing charge in the pions. For charged kaons or protons, roughly 50\% of the balancing charge will be found in the $K^+K^-$ or $p\bar{p}$ species respectively. Most of the other half of the balancing charge will be in pions. For a proton only a small fraction of the balancing charge is accounted for by kaons and vice versa.
}
\end{figure}

If $h$ and $h'$ were to refer to all charged particles, the function would integrate to unity if the measurement of $p$ were hermetic. However, even for a perfect measurement, when $h$ and $h'$ refers to a specific species pair, the integral gives, $Z\ne 1$,
\begin{eqnarray}
Z_{h|h'}&=&\int dp~B_{h|h'}(p|p').
\end{eqnarray}
This measure, or partial normalization, describes the probability of finding an excess of hadron $h$ vs $\bar{h}$ given the observation of an $h'$ hadron. Given the relative yields of the hadrons at the hyper-surface, and given the branching ratios of the decays one can calculate $Z_{h|h'}$ for all species. Ignoring decays, one can use Eq. (\ref{eq:NhofQ}) to state
\begin{eqnarray}
Z^{\rm(no~decays)}_{h|h'}&=&\int dp~B_{h|h'}(p,p')\\
\noindent
\nonumber
&=&n_h q_{hb}\chi^{-1}_{bc}q_{h'c}.
\end{eqnarray}
To include decays, one must then account for the branching ratios, $B(h\rightarrow h')$,
\begin{eqnarray}\label{eq:Zhhprime}
Z_{h|h'''}&=&
\sum_{h'h''}B(h'\rightarrow h)Z^{\rm(no~decays)}_{h'|h''}B(h''\rightarrow h''').
\end{eqnarray}
Figure \ref{fig:bfnorm} shows the partial normalization for the six combinations of species involving final-state protons, charged pions and charged kaons. Decays, including weak decays of hyperons and of neutral kaons, are incorporated. As expected, the normalization of $\pi\pi$ is nearly unity, meaning that if a charged pion is observed, there is an excess of pions of the opposite charge, versus those of the same charge, of nearly 0.9 pions; equivalently, a bit more than 10\% of the electric charge is balanced in the $p\bar{p}$ channel or in the $K^+K^-$ channel. For observation of a proton, the partial normalization due to the antiproton excess is a bit more than 0.5. The remainder of the baryon conservation would be in neutrons. Given that the proton carries both baryon number and charge, it was expected that more strength would be allotted to the proton channel, which can balance both charge and baryon number with antiprotons. Finally, for observation of a positive kaon, one finds a bit less than half of an extra negative kaon. The remainder of the balancing strangeness is in the neutral kaon channel or the hyperon channel. Because both neutral kaons and hyperons lose their strangeness in their decays it is difficult to fully account for the strangeness balance experimentally. The normalizations from Eq. \ref{eq:Zhhprime} depend on the temperature of the hydrodynamic interface. As can be seen in Fig. \ref{fig:bfnorm} the temperature dependence is modest. 

The estimates in Fig. \ref{fig:bfnorm} do not include the effects of chemical reactions during the hadronic simulation. Such reactions should not change the final-state normalizations more than a few percent. An example of such a reaction is that a $K^+K^-$ can transform into $\pi^+\pi^-$ through the $\phi$ meson channel. Baryon decays affect the width of the balance function, but have a small effect on the normalization. Thus, the partial normalizations of Fig. \ref{fig:bfnorm} should provide a good check of the integrated strengths of experimental balance functions. This will play an important role in interpreting experimental results in Sec. \ref{sec:summary}.

\section{Numerical Approach and Algorithms}\label{sec:model}

The model applied here is described in detail in Refs.~\cite{Pratt:2018ebf,Pratt:2019pnd}, where it was applied to collisions at RHIC energies and some results were compared to data from the STAR Collaboration at RHIC. A hybrid model is employed, featuring both a hydrodynamic prescription describing the high-temperature QGP phase, and a hadronic simulation for when temperatures fall below 155 MeV.  The hydrodynamic simulation was first run using the code iEBE-VISHNU \cite{Shen:2014vra}.  Parameters for the hydrodynamic code were adjusted to fit spectra and flow at the LHC. As described in Sec. \ref{sec:hydro} the evolution was based on the diffusion constant taken from lattice gauge theory \cite{Aarts:2014nba,Amato:2013naa}.  The principal difference between the calculations for this study, which are aimed at comparison with ALICE data from the LHC, and the calculations of \cite{Pratt:2018ebf,Pratt:2019pnd}, which were aimed at comparison with STAR data from RHIC, is that the ALICE data is corrected for efficiency, whereas the STAR data was not. Another small difference is that the contribution for the $\phi$ meson to the charged-kaon balance function was removed for the STAR analysis, but not for the ALICE analysis.

The charge balance functions themselves are obtained by evaluating both the numerator and denominator of each term entering the definition \eqref{Bhhp_definition}.  This requires propagating charge correlations through both the hydrodynamic and hadronic stages, as discussed above and described more fully in \cite{Pratt:2017oyf, Pratt:2018ebf}.  In this context, the numerator of each term entering the charge balance function \eqref{Bhhp_definition} receives two kinds contributions, referred to as ``Type I correlations" and ``Type II correlations." Both correlation types are calculated independently before being added together.  Here, we briefly describe both correlation types and illustrate how they are produced in our simulations.

\subsection{Type I correlations: hydrodynamic stage}
\begin{figure}
\centerline{\includegraphics[width=4.5in]{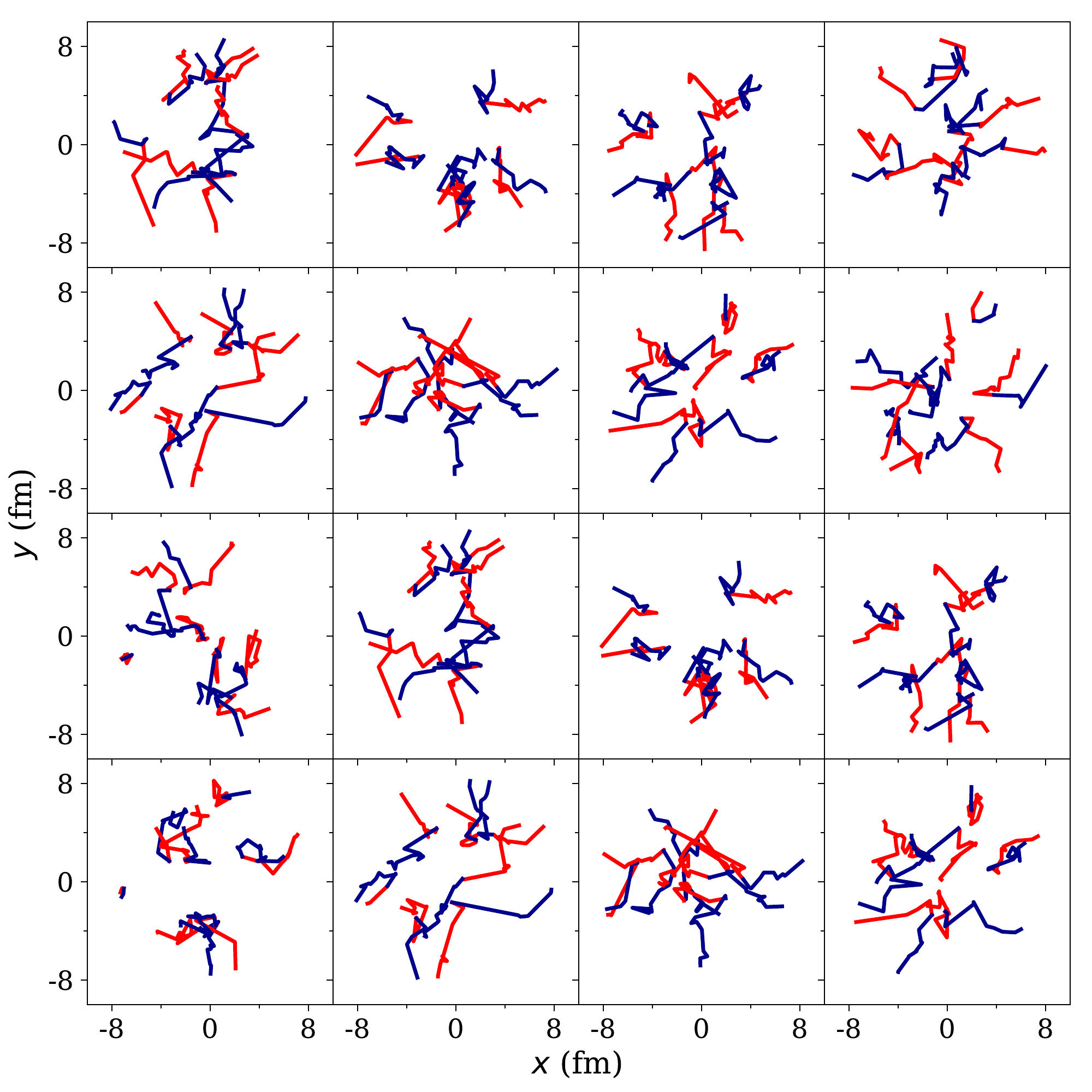}}
\caption{\label{fig:trajectories}(color online)
A sampling of trajectories of charge pairs during the hydrodynamic stage shows pairs are created at the same point, then drift apart. Collective flow focuses the trajectories of pairs to flow in similar directions. These trajectories are for the tracer particles representing the $C_{ss}$ that were created at $\tau_0$ in a calculation using the lattice values for the diffusivity. The mean typical number of collisions, or reorientations, is $\lesssim 10$. The trajectories, which show the projection of the trajectories in the transverse plane, are split into multiple panels to improve visibility.}
\end{figure}

 The first contribution comes from the hydrodynamic stage.  For these type I contribution, correlations had to be generated and evolved throughout the hydrodynamic stage. Rather than evolving $C'_{ab}(\vec{r}_1,\vec{r}_2)$ on a six-dimensional mesh, it was represented by pairs of test charges in a Monte Carlo procedure. Pairs of test charges within a space time volume $d^4x$ are generated with probability $|S_{ab}|d^4x$, where the source function is given by Eq. (\ref{eq:Sabofchi}). The charges, $dq_a$ and $dq_b$ are labeled by their charge and a weight of $\pm 1$ is assigned to the pair depending on whether $S_{ab}$ is positive or negative.  The pairs of test charges move at the speed of light in random directions, with their trajectories punctuated by ``collisions''. Initially, and after each collision, the charges have their directions randomized in the local fluid rest frame. The frequency of these collisions is set so that the ensuing random walk would reproduce the local diffusion constant in the limit of many collisions, using the relation $\tau_{\rm coll.}=6D/c^2$, with the diffusion constant being a function of the local temperature. In practice, these test charges suffer approximately ten collisions, and the representation of the diffusion equation is approximate.  Figure \ref{fig:trajectories} displays a representative sample of trajectories of test-charge pairs from the model. The projection of the trajectories in the transverse plane illustrates how charges are more likely to be focused into the same direction by the collective flow of the matter.

One difference between a full solution to the differential equation to this representation with a random-walk algorithm is that the non-causal tail of the diffusion solution is eliminated by construction, because the sampling charges do not exceed the speed of light. Once a test charge crosses into the domain of the hadron simulation it is converted into a hadron according to thermal arguments. For a differential test charge $dq_a$, when it passes through the hyper-surface and enters the domain of the hadronic simulation, the charge is converted to hadrons. For each test charge a hadron is produced with probability $\delta N_h$ as described in Eq.~\eqref{eq:NhofQ}. To increase statistics the number of generated pairs can be increased by an arbitrary factor $F$, with the incrementing of the correlation function reduced by the same factor. At the interface the correlation is expressed by summing over each pair of hadrons, where one hadron comes from the charge $dq_a$ and the other from the associated tracer charge $dq_b$. One does not consider two particles from the same tracer charge. Each pair has hadronic species $h$ and $h'$, space-time positions $x$ and $x'$, and a weight.  As pairs leave the hyper-surface, the particles are assigned a momentum consistent with the temperature and flow velocity of the local matter along with the evolution of the hyper-surface. To model the subsequent diffusion the hadrons suffer elastic collisions with a fixed cross section, colliding elastically only with a background of hadrons from simulated events. Hadrons are also allowed to decay, and all their products are included when the balance functions are constructed. This accounts for the further diffusion of particles in the hadron stage, while allowing one to track particles from a given correlated stream. The balance function is incremented according to the charges of these pairs and the pair's weight.

For each correlated hadron pair, the first particle is arbitrarily boosted to be within the acceptance to maximize statistics, while the second one is boosted by the same amount to preserve the correlation. An additional weight is then applied to the pair proportional to the width of the rapidity interval for the acceptance divided by the width of the interval in the hydrodynamic calculation. The full type I calculation required both evolving the test charges through the hydrodynamic background and evolving the converted hadrons and their daughters through the hadronic simulation. Sufficient statistics for this part of the calculation was attained with approximately 50 numerical core hours.

\subsection{Type II correlations: hadronic stage}
The second contribution to the balance function, which is labeled here as type II, comes from those correlations built up during the hadron phase. This includes contributions from two particles originating from the same decay, or from annihilations or any other chemical change. This contribution is found in a rather brute-force manner. Particles are generated for the simulation in a completely uncorrelated way, then collided just as they would for the simulation. Charge balance functions are incremented using all combinations of hadrons. This means that most of the contributions to the sums involve completely uncorrelated pairs. These sum to zero, but bring about significant noise. To overcome the noise, a total of 96,000 events were simulated, with each event covering 10 units of rapidity. Because the model was built to respect boost invariance, by applying translational boundary conditions, boosting increased the effective number of events used to sample the balance function to approximately 500,000. Even though many more events were simulated for this second contribution, it was significantly noisier than type I contributions.

Pairs that increment the balance function's numerator can be assigned weights, which may include the experimental efficiency.  ALICE's measurements accounted for efficiency, but still suffered from constraints on acceptance. Those hadrons with transverse momentum below or above the cutoffs, shown in Table \ref{table:alice_acceptance}, were discarded. Similarly, those pairs with relative rapidity greater than the maximum range for two hadrons were also discarded. For the experimental results, pairs with maximum separation in rapidity were hard to find because both particles in the pair had to be at $-Y_{\rm max}$ and $Y_{\rm max}$. Experimentally, this results in large statistical errors as $\Delta y\rightarrow 2Y_{\rm max}$. In contrast, the calculations suffered no such constraint because the translational boundaries were set at $\pm 5$ units of rapidity and cyclic boundary conditions were applied. However, the calculations could have trouble if the pairs separate by more than 5 units of spatial rapidity because they will be identified as being separated by the closest distance, so that a pair separated by 5.5 units will be assigned a relative rapidity of 4.5 units. Fortunately, very few pairs are separated by this amount as the balance functions nearly vanish by the time $\Delta y=5$.
\begin{table}
\begin{tabular}{|c|c|c|c|}
\hline
species & rapidity & tranverse momentum & Distance of Closest Approach\\
\hline
$\pi\pi$ & $|y|<0.8$ & $0.2<p_t<2.0$ GeV/$c$ & $DCA_z<2.0$ cm, $DCA_{xy}<0.04$ cm\\
$KK$ & $|y|<0.7$ & $0.2<p_t<2.0$ GeV/$c$ &  $DCA_z<2.0$ cm, $DCA_{xy}<2.0$ cm\\
\hline\end{tabular}
\caption{\label{table:alice_acceptance}Acceptance of ALICE analysis. $\sqrt{s_{NN}}=2.76$ TeV, $z-$vertex $\le 6$ cm, $10^7$ collisions over all centralities.
}
\end{table}

\begin{figure}
\centerline{\includegraphics[width=0.6\textwidth]{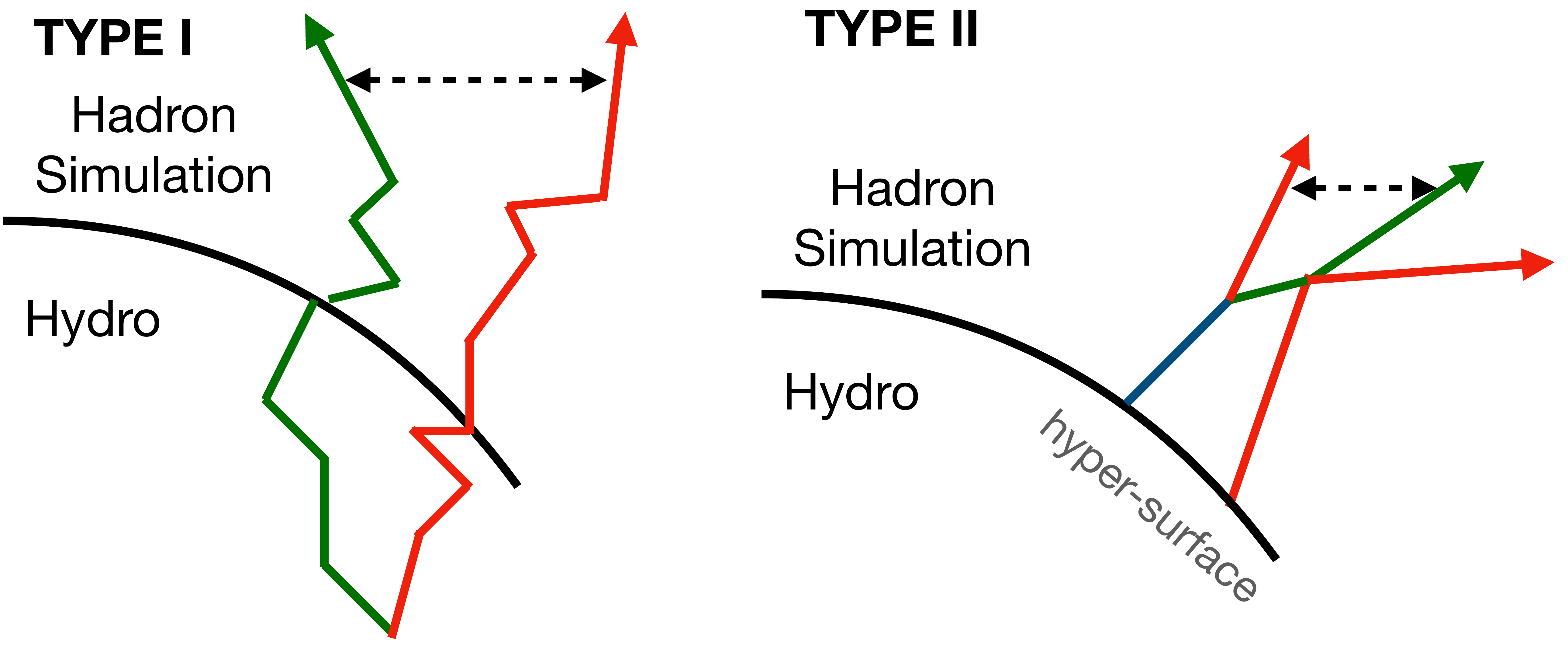}}
\caption{\label{fig:type1type2}(color online)
An illustration of the two contributions to the balance function calculations. (Left) Type-I contributions derive from correlations that have been evolved through the hydrodynamic stage, where they are represented by tracer charges. Tracer particles are then converted with statistical weights to hadrons and are followed through a simulation of their collisions and decays. The balance function numerators are then incremented by combining hadrons from each of the tracer charges. No contributions are generated from hadrons who derive from uncorrelated pairs of tracer charges, and pair from the same tracer charge are also neglected. (Right) Type-II contributions are generated by simply generating uncorrelated particles from the hydrodynamic/simulation hyper-surface, then combining all hadrons afterwards. These correlations are mainly those from decays, and by using a simulation accounts for the scattering of the decay products. By considering all pairs, similar to what is done with experiment, the contribution from type-II have significant statistical error.}
\end{figure}

Finally, the two types of contributions simply add when calculating the balance function. Figure \ref{fig:type1type2} shows how each type contributes to the balance function for each of the six species-dependent balance functions. As expected, the type-I contributions tend to be broader in rapidity, while the type-II contributions are all narrow. Each contribution to $B_{h|h'}$ from the type-I contributions can be traced back to the source function at a particular point in space-time \cite{Pratt:2017oyf}. From Fig.~\ref{fig:chi} one can see that significant type-I contributions derive from the changing susceptibilities near, but still above, the interface temperature. This is especially true for the off-diagonal terms for $\chi_{ab}$. For example, the off-diagonal term $\chi_{us}=\chi_{ds}$ provides the dominant source for the $Kp$ balance functions \cite{Pratt:2017oyf}.

\section{Results: Sensitivity to Initial Separation\label{sec:sigma}}

In this section, we consider the sensitivity of our model predictions to the initial separation between charge pairs at the beginning of the hydrodynamic phase.  We focus on determining which separations appear to be favored by the ALICE balance function measurements. Only the most central collisions are considered in this study, i.e. the 0-5\% most central collisions.

For charges emitted from the same point, their separation in rapidity is determined by their mass and temperature. The variance of the rapidity relative to the spatial rapidity for a single particle is approximately $T_b/M_\perp $, where $M_\perp$ is the transverse mass and $T_b$ is the temperature at breakup. Heavier particles, like protons, have lower thermal velocities and are thus would be more highly correlated in rapidity. Two more parts of the physics affect the width of the balance function, $B(\Delta y)$. First, there is a separation due to the fact that at $\tau_0$ particles may already have moved from the point at which a pair originated. To account for how far a particle has moved from the point at which a pair was created to its position at the time, $\tau_0$, when hydrodynamics begins, each tracer charge has moved in spatial rapidity according to a Gaussian distribution with variance $\sigma_0$. Thus, the initial separation in spatial rapidity between two particles would be described by a Gaussian with variance $2\sigma_0^2$. The median separation between two balancing charges at $\tau_0$ is slightly less than $\sigma_0$. The tunneling involved in the dissolution of chromo-electric flux tubes might contribute to the width $\sigma_0$, or the width might be due to charges being created at times less than $\tau_0$ followed by some spreading. From charge balance functions measured in $pp$ collisions or in peripheral heavy-ion collisions one might expect $\sigma_0$ to be on the order of a half unit of rapidity. The third contribution comes from the diffusion of the charge and its balancing charge between $\tau_0$ and the final time $\tau_f$. As time increases the diffusion constant should increase as the density falls and the mean free path increases, and if the cross section is fixed and if the thermal velocities are fixed, which would be true for massless particles, one would expect $D(\tau)$ to increase linearly with time. Assuming $D=\beta\tau$, the diffusive separation then increases logarithmically with time \cite{Bass:2000az},
\begin{eqnarray}
\sigma_y^2=2\sigma_0^2+2T_b/M_\perp +4\beta\ln(\tau_f/\tau_0).
\end{eqnarray}
This expression grossly over-simplifies the physics, but it is useful in that it emphasizes that diffusion at early times plays an outsized role in the final width of the balance function in relative rapidity. For example, the separation due to diffusion of two charges between $\tau_0$ and $2\tau_0$ plays as important a role as diffusion between $\tau=5$ fm/$c$ and $\tau=10$ fm/$c$. For the detailed model presented here the structure of the balance function is driven by the same factors. First, there is the initial separation of the charges at the time of thermalization, $\sigma_0$. The diffusivity, the time from when charges are created until breakup, and the final breakup temperature, all affect the width. The principal goal of this section is to understand the sensitivity to $\sigma_0$.

For the calculations presented here, evolution begins at the time $\tau_0=0.6$ fm/$c$. At such an early time charges may have already separated by a few tenths of a femtometer. In a central collision such a small separation is negligible in regards to the relative transverse diffusive separation because the overall transverse size is $\sim 5$ fm, and adding few tenths of a fm in quadrature would have little effect. However, in the longitudinal direction such a separation can have a large effect due to the large initial longitudinal flow. The difference in spatial rapidity is $\Delta\eta_s\approx\delta z/\tau$, so a 0.3 fm separation in coordinate space translates to a half unit of rapidity, which is significant as the separation will be magnified by longitudinal collective flow. Thus, the parameter $\sigma_0$ clearly affects the widths of the charge balance functions when binned by rapidity. The angle-binned balance functions are also affected, but mainly because the normalization of $B(\Delta\phi)$
is reduced for larger $\sigma_0$ because it becomes less likely that a charge and its balancing charge will both fit in the rapidity window.

\begin{figure}
\centerline{\includegraphics[width=0.8\textwidth]{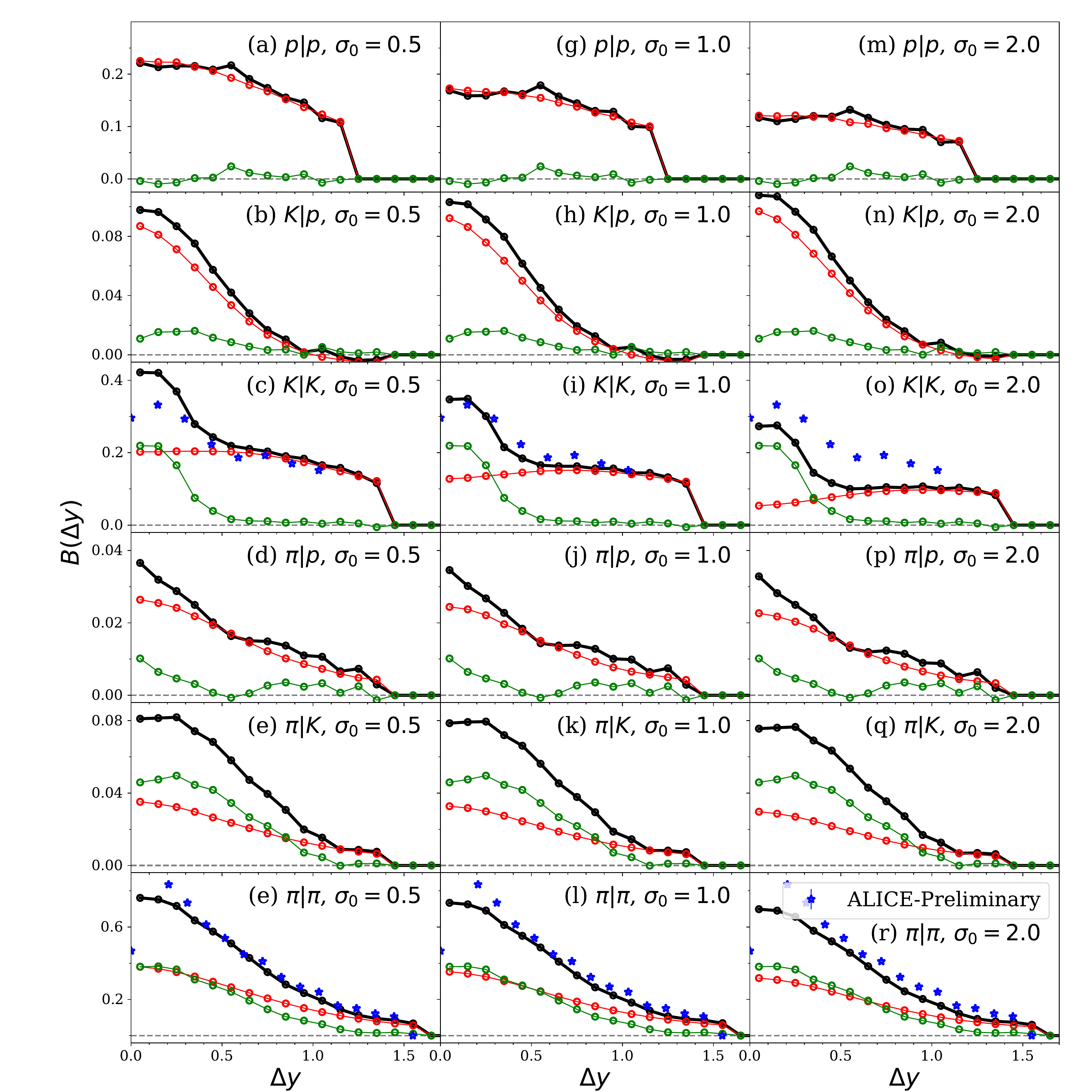}}
\caption{\label{fig:general_y_varysigma}(color online)
Balance functions indexed by species and binned by relative rapidity are displayed alongside preliminary experimental results from ALICE. Type-I contributions (red circles) and type-II contributions (green circles) are summed to construct the correlation (black circles). Early creation of the QGP was expected to result in broader balance functions for kaons and protons than for pions.  Indeed, in the model calculations these balance functions were found to be broader than the pion balance functions despite the fact that thermal motion  more broadly spreads the charge balance for pions than for kaons or protons, which have less thermal velocity due to their greater masses. Experimental results from ALICE  are in line with model calculations, both qualitatively and quantitatively. Balance functions binned by rapidity are sensitive to $\sigma_0$, which sets the distribution of relative spatial rapidities between balancing charges when the hydrodynamic calculation is instantiated at $\tau_0=0.6$ fm/$c$. Calculations seem to have a preference for $0.5<\sigma_0< 1$.}
\end{figure}

Here, we compare the full model to ALICE results. Calculations employed the diffusion constant, $D(T)$, from lattice calculations \cite{Aarts:2014nba,Amato:2013naa}. The temperature dependence of the susceptibility, $\chi_{ab}(T)$, and the equation of state driving the hydrodynamic acceleration were also taken from lattice calculations \cite{Borsanyi:2011sw}. The initial width $\sigma_0$ is not constrained by lattice calculations. It varies the width of the charge-charge correlation functions in spatial rapidity at $\tau_0$,
\begin{eqnarray}
C_{ab}(\Delta\eta_s,\tau_0)&\sim e^{-(\Delta\eta_s)^2/4\sigma_0^2}.
\end{eqnarray}
Because the mechanism and time scale of initial charge production is not well known, especially in the context of a central heavy-ion collision, this parameter might be on the order $\lesssim 1$ units of spatial rapidity, but there is no good experimental evidence to constrain it tightly. Figure \ref{fig:general_y_varysigma} shows balance functions for all six combinations of $\pi$, $K$ and $p$. Balance functions are filtered through the ALICE acceptance. Unlike STAR analyses from RHIC, these have been corrected for efficiency and acceptance. They are constrained to a range in $\Delta y$ and by the transverse momentum of particles. Table \ref{table:alice_acceptance} shows the range of the ALICE acceptance.

As expected, given the behavior of the susceptibilities in Fig. \ref{fig:chi}, the $\pi\pi$ balance functions are narrower than either the $pp$ or $KK$ balance functions. The off-diagonal susceptibilities, which become non-zero only when the matter cools to the hadronization region, also contribute to the narrow structures, particularly to that of the $Kp$ balance function. The agreement between model and experiment makes a strong case that the matter created in central collisions at the LHC approaches chemical equilibrium at times $\lesssim 1$ fm/$c$. If the matter were to spend several fm/$c$ as a gluon plasma, with quarks only gradually appearing, the charge balance functions would be narrower, especially for $pp$ and $KK$. If all charges were created close to hadronization, the $\pi\pi$ balance function would be broader than the $KK$ or $pp$ balance functions. 

\begin{figure}
\centerline{\includegraphics[width=0.8\textwidth]{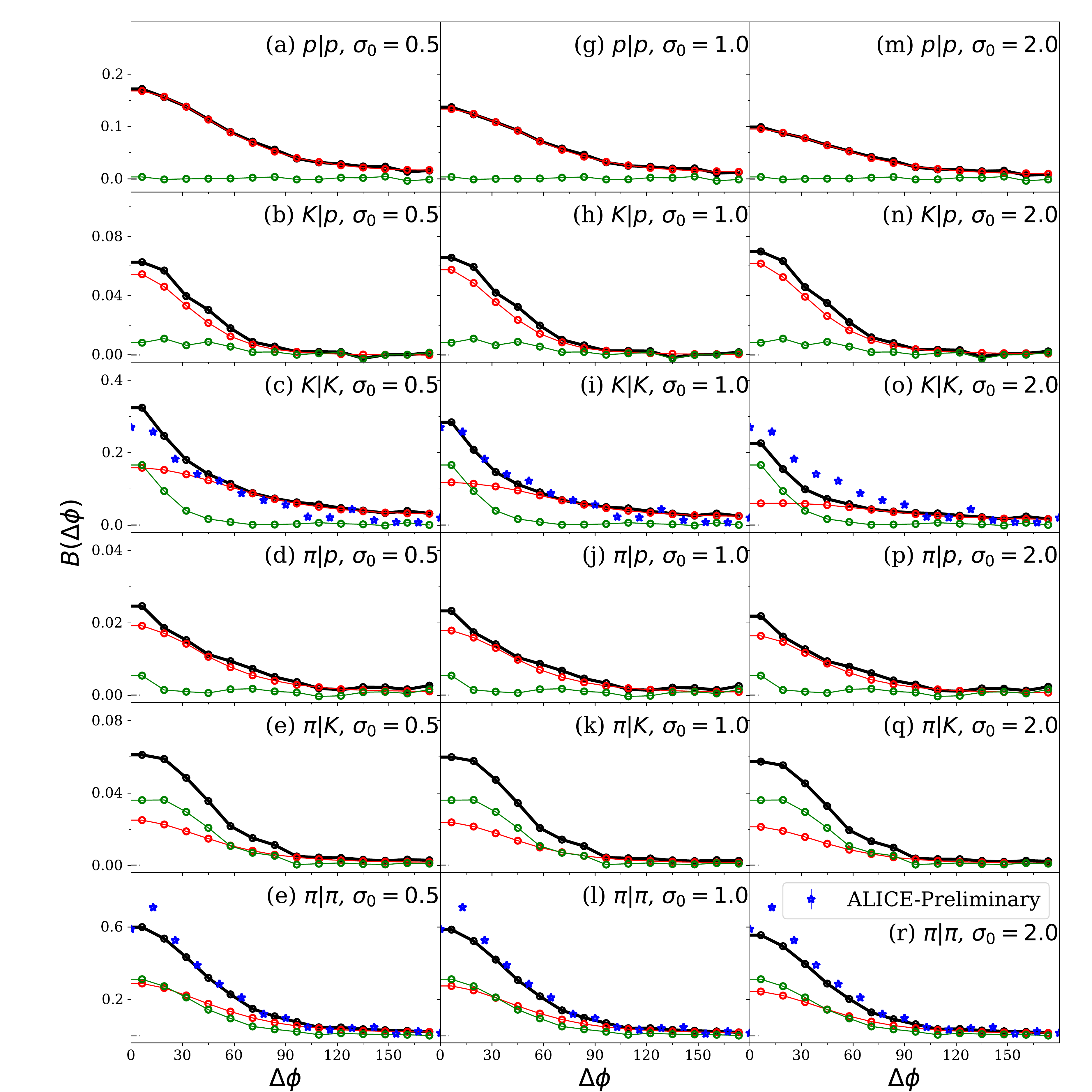}}
\caption{\label{fig:general_phi_varysigma}(color online)
Balance functions binned by relative angle are shown for several values of $\sigma_0$, the initial spread in spatial rapidity of balancing charges at $\tau_0$, the time hydrodynamics is instantiated. Larger values of $\sigma_0$ reduce the probability of balancing charges being recorded within the acceptance. In turn, this mainly lowers the balance functions binned by relative azimuthal angle because more balancing charges are not found within the rapidity acceptance. As was the case for balance functions binned by $\Delta y$, the best fit with preliminary ALICE data appears to be for $0.5<\sigma_0<1$. Type-I contributions (red circles) and type-II contributions (green circles) are summed to construct the correlation (black circles).}
\end{figure}

Figure \ref{fig:general_phi_varysigma} displays the sensitivity of charge balance functions to the parameter $\sigma_0$. Even though the width of the balance functions in $\Delta\phi$ is only modestly affected, the normalization noticeably changes. This was expected, because for larger values of $\sigma_0$ more of the charge balance moves to larger values of $\Delta y$ which are outside the acceptance. 

If $0.5<\sigma<1.0$, as suggested by the data comparison in Figs. \ref{fig:general_y_varysigma} and \ref{fig:general_phi_varysigma} it would have significant implications for understanding the thermalization stage of the QGP. For example, for $\sigma_0=0.75$ half the original balancing pairs are separated by more than 0.7 units of spatial rapidity at $\tau_0$. This could be explained by either having  charges created significantly before $\tau_0$ or if quark-antiquark pairs were created by the tunneling process of longitudinal flux tubes dissociating. The latter would not be possible if a gluon plasma were to thermalize and become quasi-isotropic before quarks appear, because the pairs would not likely be preferentially separated along the $z$-axis. The role of jets, or mini-jets, should be analyzed in greater detail. Gluon jets that dissociate into quarks should result in highly correlated quarks, though if the dissociation occurs early the quarks would separate diffusively. Thus, quarks production from gluon jets should not be much different than those from flux tubes. The effect of quark jets might be quite different, because such quarks might be separated by large ranges in rapidity. Even though one expects the percentage of quarks from quark jets to be small, they might provide a significant fraction of the balance function's strength at high relative rapidity. This possibility needs to be studied in greater detail. Finally, extracting the centrality dependence of $\sigma_0$ would help determine whether early stage thermalization changes character for central events. It does seem clear that quark production occurs early, on the scale of 0.5 fm/$c$ or earlier, but it is premature to make strong conclusions about the mechanism for quark production at the current time. The sensitivity to $\sigma_0$ is modest and experimental results have significant uncertainties. 

The ATLAS and CMS Collaborations have the ability to better measure charge balance functions at larger relative rapidity. Even if such measurements would not allow particle identification, the balance functions of non-identified (aside from charge) particles for relative pseudo-rapidities $\gtrsim 2.0$ would be enlightening. Looking forward, the CMS Collaboration has plans implement an upgrade that would enable particle identification up to pseudo-rapidities near 3.0 \cite{Lucchini:2020iea}, which would roughly triple the rapidity range of the ALICE analysis. For heavier particles, especially those at lower $p_t$, the range in real rapidity is significantly reduced from the pseudo-rapidity range, so the CMS upgrade would be especially useful for proton and kaon balance functions. However, future experimental analyses should have much better statistics, and if the acceptance in rapidity can be increased, balance function measurements might be able to address the mystery of how quark production takes place, and how chemical equilibrium might be attained at such short times.

Charge annihilation at later stages of the collision dampens the balance function at small relative rapidity, and because the normalization is nearly fixed by conservation constraints, the width would increased. Femtoscopic correlations, which were crudely studied in \cite{Jeon:2001ue} might also affect the answer non-negligibly. These latter two caveats can be better understood by more careful modeling, and greatly improved experimental statistics should be analyzed soon.

\section{Balance Functions and Diffusivity}\label{sec:diffusion}

Here, we continue the discussion of our model results and explore the sensitivity of these results to the strength of diffusivity employed in the hydrodynamic phase.  Again, we consider the implications of this sensitivity for the interpretation of the relevant ALICE data.

The diffusivity can be most effectively probed by studying the balance functions in terms of their dependence on the pair separation in azimuthal angle.  Collective radial flow focuses balancing charges into the same direction if the balancing charges are close to one another, but less so if they have had the opportunity to drift apart. Due to the strong longitudinal flow at early times, $v_z\approx z/t$, balance functions binned by relative rapidity are sensitive to the initial correlations at $\tau_0$.  Balancing charges separated by 0.6 fm in the longitudinal direction at $\tau_0=0.6$ fm/$c$ are already separated by a unit of spatial rapidity, and collective flow will pull them apart by $\approx 20$ fm by $\tau=20$ fm/$c$ even if they do not diffuse further apart. In contrast, the transverse separation of charges is not much changed if they are already separated by a few tenths of a fm at $\tau_0$ because transverse flow is small at early times, at least for central collisions. The parameter $\sigma_0$ has little effect on how charges are separated transversely, whereas it significantly affects the width of the balance function binned by relative rapidity. Thus, balance functions binned by relative azimuthal angle, $\Delta\phi$, provide a clearer means for determining the diffusivity of the matter. To determine the diffusion constant, one needs to know both the final separation of the charges, and the time at which they were created. For this reason $pp$ and $KK$ charge balance functions are especially useful for constraining the diffusivity because the source functions which drive them are almost entirely concentrated at the earliest times.

\begin{figure}
\centerline{\includegraphics[width=\textwidth]{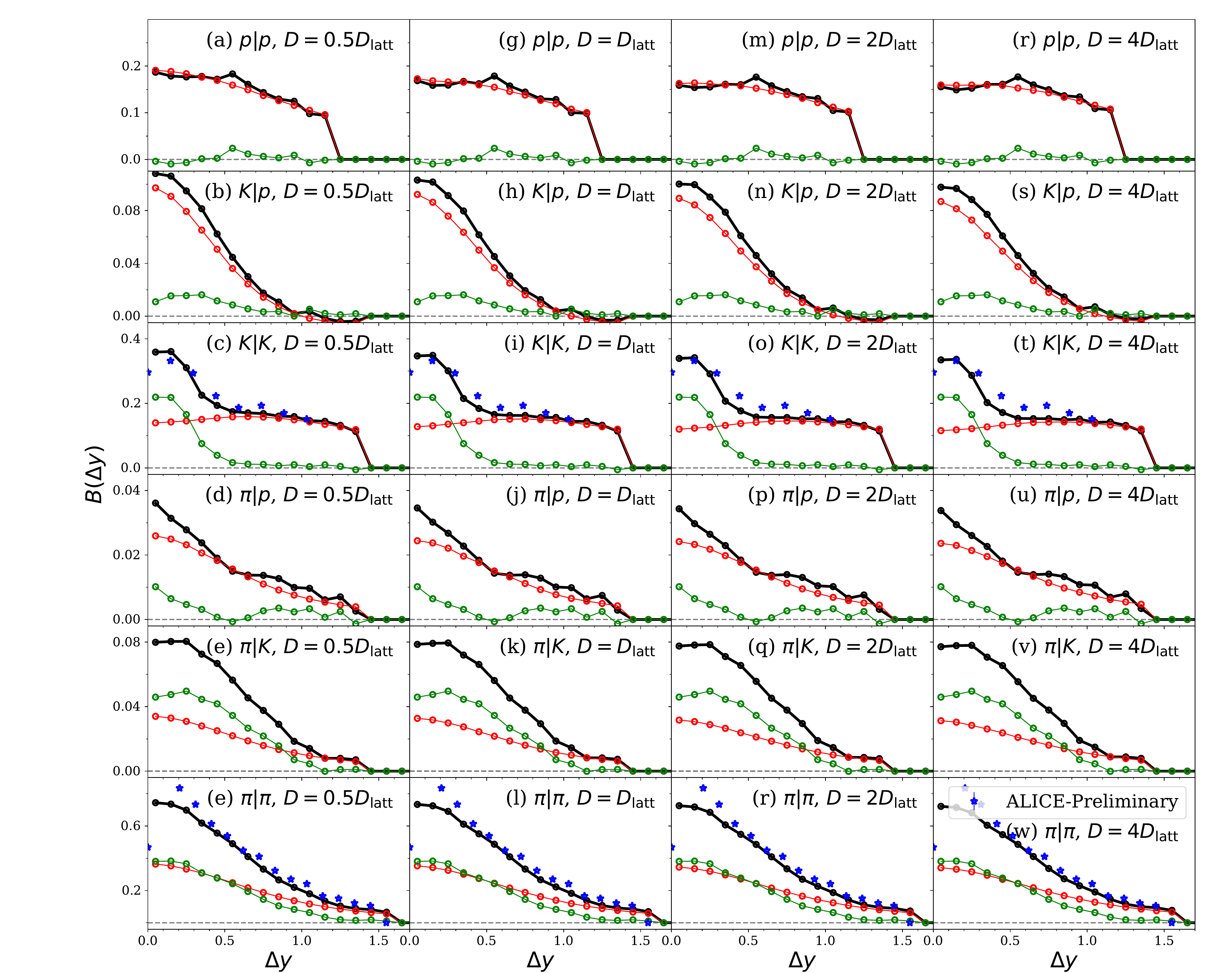}}
\caption{\label{fig:general_y_varyD}(color online)
Balance functions, binned by relative rapidity are displayed for different diffusivities. Larger diffusivities have the same effect as increasing the value of $\sigma_0$ as shown in the previous section. The $KK$ and $pp$ balance functions are more sensitive to the diffusivity because they owe more of their strength to type-I contributions (red circles) than to type-II contributions (green circles), those correlations given birth at the earliest times, or during the hydrodynamic phase.}
\end{figure}

\begin{figure}
\centerline{\includegraphics[width=\textwidth]{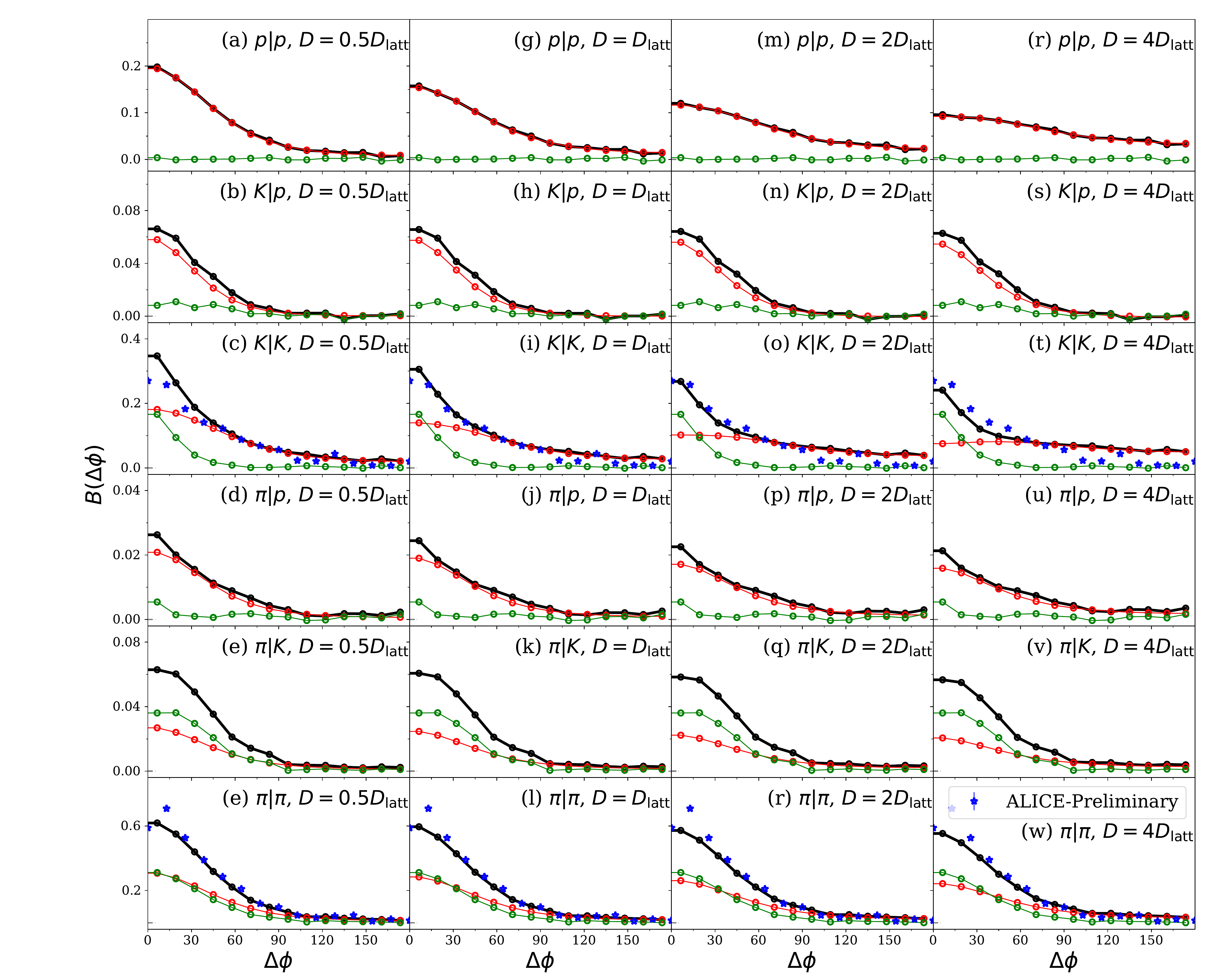}}
\caption{\label{fig:general_phi_varyD}
Balance functions binned by $\Delta\phi$ are broader for higher values of the diffusivity. Whereas the height of these same balance functions were sensitive to $\sigma_0$, the width is mainly driven by the diffusivity. This is especially true for the $KK$ and $pp$ balance functions, which are mainly sourced at early times. Comparison with preliminary ALICE results suggests that diffusivities close to those used for lattice gauge theory are consistent with data, and that doubling or quadrupling the diffusivity leads to somewhat less satisfactory reproductions of experimental results. Type-I contributions (red circles) and type-II contributions (green circles) are summed to construct the correlation (black circles).}
\end{figure}

The sensitivity to the diffusivity is shown in Figs.~\ref{fig:general_y_varyD} and \ref{fig:general_phi_varyD}. In the base calculation the diffusivity was set as a function of temperature according to lattice results \cite{Aarts:2014nba,Amato:2013naa}. As expected, the $KK$ and $Kp$ balance functions are significantly sensitive to the diffusivity. Doubling the diffusivity can affect the balance function by several tens of percent, which makes one optimistic about the prospects of extracting the diffusivity from experiment. Comparison with data in Fig.~\ref{fig:general_phi_varyD} shows that the diffusivity from lattice calculations appear remarkably consistent with measurements from ALICE. Calculations with half or double the diffusivity seem less able to reproduce ALICE measurements, but conclusions must be tempered as discussed in the following paragraphs. 

It should be emphasized that the behavior at small $\Delta y$ or $\Delta\phi$ can be easily altered by the chemical evolution in the hadronic phase. Baryon-antibaryon annihilation \cite{Pan:2014caa,Steinheimer:2017vju} was not included in the calculations, but could easily suppress the $pp$ balance function by tens of percent near $\Delta\phi=0$. This suppression would then increase the strength at larger $\Delta\phi$, or at larger $\Delta y$, because it also lowers the denominator is such a way as to maintain the charge conservation constraints. Thus, increasing the diffusivity and introducing baryon annihlation can have similar effects. Strangeness can also annihilate. For example the reaction $K^+K^-\rightarrow \phi\rightarrow \pi\pi$ can have the same effect on the $KK$ balance function as baryon annihilation does for the $pp$ case. Such effects are probably rather small, but nonetheless this introduces uncertainty into any inference of the diffusivity from the $B_{K|K}(\Delta\phi)$. Another class of effects that alters balance functions at small $\Delta\phi$ is final-state interaction (FSI) between the emitted particles. Identical-particle interference and Coulomb interaction drive correlations at small relative momentum that provide the means to femtoscopically extract source size and lifetime information. However, for these analyses these effects are ignored. Again, by depressing or suppressing the balance functions at small $\Delta\phi$, they must also enhance the strength at larger $\Delta\phi$ because the sum rules still apply. In \cite{Jeon:2001ue} these effects were found to be small, but non-negligible, for balance functions binned by relative rapidity, whereas the balance function binned by $\Delta\phi$ play the critical role here. 

Even $pp$ and $KK$ balance functions have contributions from later times. For example, $\phi$ meson decays contribute to a narrow peak in the $KK$ balance function and $pp$ balance functions can have dips at small relative momentum due to annihilation. One can avoid these complicating factors by focusing on balance functions binned by $\Delta\phi$ while also considering only pairs with larger relative rapidities. As was seen in the previous sections, the physics of the hadronization stage has negligible effect on the numerators of $pp$ and $KK$ balance functions for $\Delta y\gtrsim 1$ units of rapidity. Because balance functions have constrained normalizations, the physics of the hadronization phase can affect the denominator. For example if 25\% of the protons annihilate with antiprotons, the numerator of the balance function is reduced by 25\% and the balance function at larger $\Delta y$ increases by a factor of $4/3$. To avoid this effect, one can consider the ratio of the balance function binned by $\Delta\phi$ to one integrated over $\Delta\phi$. In particular, we propose the following observable which should be mostly insensitive to such effects:
\begin{eqnarray}\label{eq:R1def}
R_1(\Delta y)&=&\frac{B_1(\Delta y)}{B(\Delta y)}\equiv\frac{\int d\Delta\phi~B(\Delta y,\Delta\phi)\cos(\Delta\phi)}{B(\Delta y)}. 
\end{eqnarray}
The denominator of the balance function disappears when constructing this ratio and one is sensitive only to the correlation at a specific $\Delta y$. By avoiding $\Delta y< 1$ unit, one can isolate those charge pairs which were created at early times, especially in the $KK$ and $pp$ balance functions. Additionally, this ratio should be independent of $\sigma_0$ as a balancing pair's relative angle should be independent of how far apart they were created in spatial rapidity, as long one knows the pair was created early. Thus, this ratio should provide nearly unambiguous insight into the diffusivity.

\begin{figure}
\centerline{\includegraphics[width=0.6\textwidth]{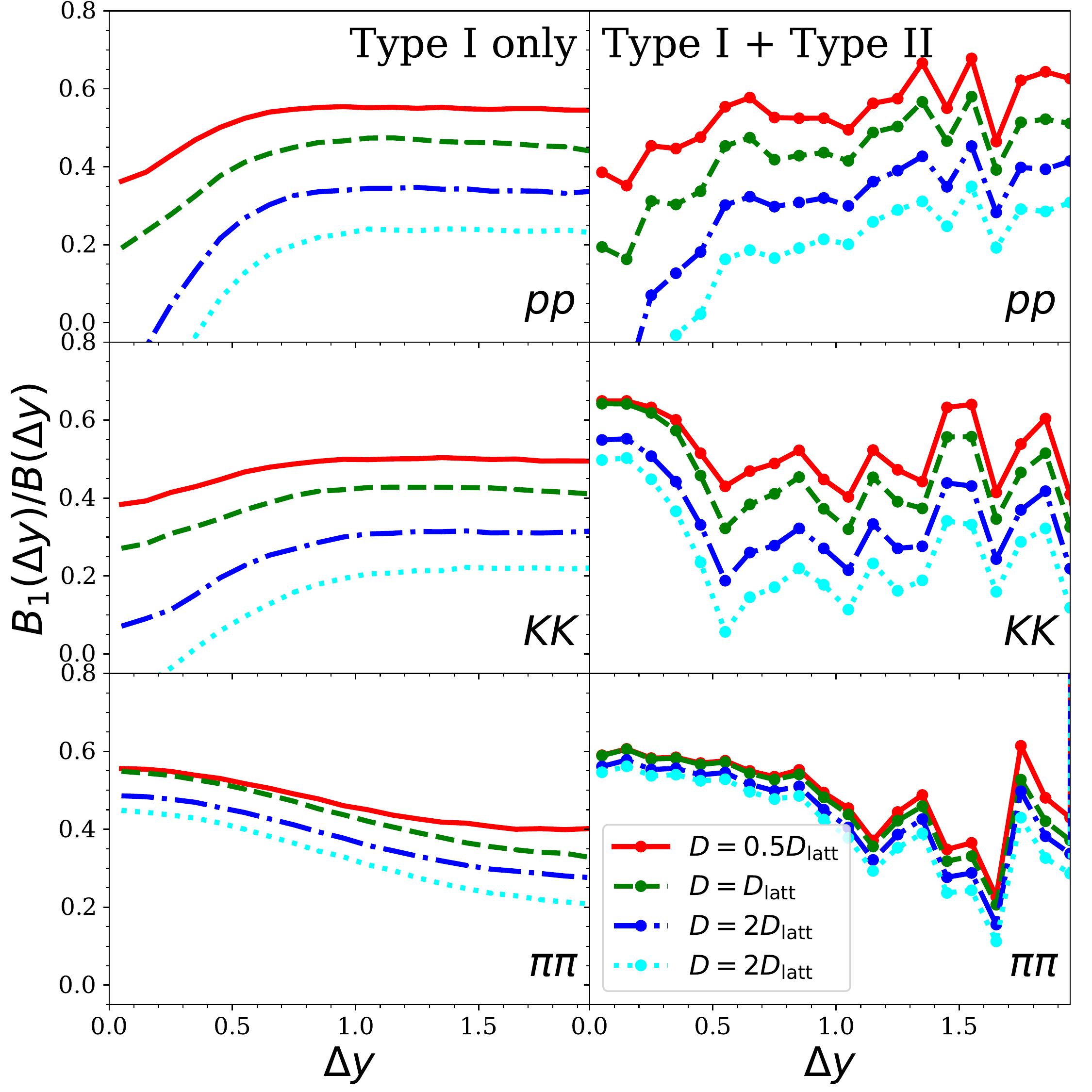}}
\caption{\label{fig:bf_y1}(color online)
The ratio $R_1(\Delta y)$, defined in Eq. (\ref{eq:R1def}) provides strong resolving power for describing the viscosity. This is especially true for the $KK$ and $pp$ constructions. For $\Delta y\gtrsim 1$ balance functions are dominated by type-I contributions. Due to the fact that $B(\Delta y)$ is becoming small for larger $\Delta y$, and because the calculation of type-II contributions is relatively noisy, even when the contribution is small, the calculations for $R_1$ become increasingly noisy when both types are included in the calculation. However, the type-I contributions by themselves (left panels) should accurately represent $R_1$ for $\Delta y>1$.}
\end{figure}

Figure \ref{fig:bf_y1} shows $R_1(\Delta y)$ for $\pi\pi$, $KK$ and $pp$ balance functions for several diffusivities. Calculations were performed with an acceptance that is independent of acceptance so as to explore the behavior for rapidities beyond what ALICE can measure. The only acceptance cuts were that transverse momenta were required to exceed 300 MeV/$c$. As expected from the discussion above the ratio becomes independent of $\Delta y$ for both $KK$ and $pp$ balance functions. The ratio is shown for the type-I contributions on the left and for the entire balance function (i.e., both contribution types) on the right. For $\Delta y\gtrsim 1$ the contribution to $B(\Delta y)$ from the type-II correlations should be negligible. However, because the type-II correlations are fraught with noise, and because one is dividing by $B(\Delta y)$ in a region where $B(\Delta y)$ is small, the noise from the type-II calculation significantly contributes to $R_1(\Delta y)$ for large regions of $\Delta y$, where the type-II contributions should be negligible. Thus, the ratio without the type-II correlations is shown alongside to better show what the ratio should be for $\Delta y\gtrsim 1$.

The separation between various diffusivities in Fig. \ref{fig:bf_y1} is striking. Doubling the diffusivity from the lattice values decreased the asymptotic value of $R_1$ by roughly 30\%. If measurements can be performed with sufficient statistics, and if theoretical uncertainties can be reduced to the 10\% level, one should be able to constrain the diffusivity to better than 50\%. This might be somewhat better than the degree to which the shear viscosity is currently being extracted from experiment \cite{Sangaline:2015isa,Bernhard:2019bmu}. The type-II contributions were calculated using statistics comparable to an experiment with roughly 500,000 events in a given centrality. In comparison, the preliminary ALICE measurements were performed using $10^7$ events of all centralities, so the noisy contributions from the type-II calculations in Fig. \ref{fig:bf_y1} are probably fairly indicative of the level of noise the ALICE collaboration would encounter should they produce the corresponding plot from data, though the range in $\Delta y$ would be abbreviated. Fortunately, data sets from ALICE with much higher statistics should become available in the coming years. Also, the same ratio can be constructed using unidentified particles binned by pseudo-rapidity. This suggests that CMS and ATLAS, with their extended acceptance in rapidity but with a lack of particle identification, may be able to construct this observable out to high values of relative pseudorapidity. If the CMS upgrade \cite{Lucchini:2020iea} is implemented the ability to analyze the balance functions at large relative rapidity would be greatly enhanced.

\section{Summary}\label{sec:summary}

The model applied here was the same as what has been used to address previous data from RHIC. The only significant modifications were that the initial hydrodynamic conditions were adjusted for the higher energy densities of the LHC. By comparing to preliminary data from the ALICE Collaboration two fundamental questions were addressed. The first question concerns whether a chemically equilibrated QGP was created in collisions at the LHC, and the second is whether the diffusivity of the matter in the super-hadronic phase can be constrained. Both questions are difficult to address with any other class of measurements. The model-data comparisons presented here suggest that the answer to both of these questions is ``yes'', and further that the diffusivity is not far from predictions of lattice calculations. Two model parameters were adjusted in this study. The first is $\sigma_0$, the width in relative rapidity of the charge balance at the time $\tau_0=0.6$ fm/$c$, when the hydrodynamic calculation was begun. The second model variation was in adjusting the diffusivity. Compared to a default calculation using $D_{ab}(T)$ from lattice calculations, the diffusivity was varied by re-running the calculation with various multiples of $D(T)$. The resolving power of the experimental data strongly rests on being able to consider charge balance functions indexed by hadron species, especially by using the $\pi\pi$, $pp$ and $KK$ balance functions. This had been done at RHIC for balance functions binned by relative rapidity, $\Delta y$, but not for those binned by relative azimuthal angle, $\Delta\phi$. Here, ALICE's measurement of $B_{K|K}(\Delta\phi)$ proved especially important in addressing both questions posed above. 

The question of whether a chemically equilibrated QGP had already been created at the time $\tau_0$, two pieces of experimental information were particularly important. First, the width of $B_{K|K}(\Delta y)$ was noticeably larger than that of $B_{\pi|\pi}(\Delta y)$. Because the ALICE Collaboration was able to correct their results for acceptance, this feature was more apparent in ALICE data than it was in STAR data. Broad $KK$ balance functions are expected only if balancing strange-antistrange charges are produced earlier in the collision. In contrast, $B_{\pi|\pi}(\Delta y)$ is largely driven by the large surge of up and down quark production that occurs during hadronization and in the decays of heavy resonances, and the width of the $\pi\pi$ balance function is only modestly sensitive to the initial chemistry.  Not only were the behaviors of the $\pi\pi$ and $KK$ balance functions qualitatively consistent with expectations, they were quantitatively consistent as long as $\sigma_0$ was in the neighborhood of $0.75$. For such a width to be attained at $\tau_0=0.6$ fm/$c$, one would infer that most of the charge was either created before $\tau_0$, or if created at $\tau_0$ the charge creation mechanism involved something like the tunneling associated with breaking longitudinal flux tubes, which could pull charges apart so that the field energy could be transformed into the quark energies. Unless the tunneling mechanism separated charges by distances $\gtrsim 0.5$ fm, which seems rather unlikely, such a mechanism must have large finished by $\tau_0$ if the separation is large enough to approach the value of $\sigma_0$ extracted here. Thus, it is difficult to imagine how any scenario where quark production is delayed beyond $\tau_0=0.6$ fm/$c$ could lead to the broad $KK$ balance functions found here. This same evidence had been seen in STAR results, but conclusions from RHIC data had been somewhat more guarded because the diffusivity, which also plays a role in the width of $B_{K|K}(\Delta y)$, was not as well constrained.

The width of the $KK$ or $pp$ balance functions in azimuthal angle plays a critical role in constraining the diffusivity. Both balance functions are primarily sourced by the creation of the initial correlation at $\tau_0$. Calculations using the lattice diffusivity were remarkably able to reproduce ALICE measurements of both $B_{K|K}(\Delta\phi)$ and $B_{\pi|\pi}(\Delta\phi)$, where as calculations using half or twice the lattice diffusivity were somewhat less successful, and calculations with quadruple the lattice diffusivity noticeably failed to reproduce the data. As expected, most of the sensitivity to the diffusivity came from the $KK$ balance function. However, conclusions are muted somewhat by realizing that the calculations might have missed some physics that adjust the correlations at small relative momentum. This includes annihilation in the hadronic phase, perhaps missing some resonances, or correlations from final-state interactions. To avoid these caveats calculations were presented of a novel observables, the ratio $R_1(\Delta y)$ defined in Eq. (\ref{eq:R1def}). For a given $\Delta y$ bin, $R_1$ provides a measure of the width of the balance function in $\Delta\phi$, and avoids any physics that might change the magnitude of $B(\Delta y)$. For larger values of $\Delta y$, this provides a more robust means for constraining the diffusivity. This can be accomplished if experiments can generate sufficient statistics to extract $R_{1,pp}(\Delta y)$ for $\Delta y\gtrsim 0.7$, $R_{1,KK}(\Delta y)$ for $\Delta y\gtrsim 0.9$, or $R_{1,\pi\pi}(\Delta y)$ for $\Delta y\gtrsim 1.5$. The sensitivity of $R_1(\Delta y)$ to the diffusivity as displayed in Fig. \ref{fig:bf_y1} is reason for great optimism. 

Both experiment and theory can improve in ways that should significantly clarify the answers to these questions. For experiments, both at the LHC and at RHIC, the greatest need is to increase the statistics of the analysis. The preliminary ALICE analysis used to generate the balance functions here used $10^7$ events but these events covered all centralities. This is similar to the statistical resolving power of the type-II calculations used here, and the noise in model calculations for Fig.~\ref{fig:bf_y1} suggests that a ten-fold increase in statistics would result in statistical fluctuations of $R_1(\Delta y\approx 1)$ significantly below the variations of $R_1$ due to doubling or halving the diffusivity. Aside from improving statistics, experiments should also present $pp$ balance functions alongside the $KK$ and $\pi\pi$ results shown here. Although the ALICE detector identifies particles with relative pseudo-rapidities up to 1.6, the range in real rapidity is signficantly lower, especially for protons. The planned upgrade of the CMS detector \cite{Lucchini:2020iea} would greatly enhance measurements of the balance function at larger relative rapidity, and thus much better address the mysteries of quark production at early times and to better constrain the diffusivity. For theory, several aspects of the model calculations stand out as needing attention. First, the role of annihilation in the hadron phase, especially of baryons, should be added. This should have relatively little effect on $B_{K|K}$ and almost no effect on $R_1(\Delta y\approx 1)$, but could significantly affect the other balance function observables, especially those involving the $pp$ balance functions. Second, the role of final-state interactions needs to be better quantified. A crude estimate exists for how $B_{\pi|\pi}(\Delta y)$ is affected \cite{Pratt:2003gh}. However, effects could vary by species. Third, if one is to analyze balance functions for less central collisions, the initial separation of the balancing charges in the transverse directions might become important. The model used here would need to introduce some uncertainty for this separation similarly to how the parameter $\sigma_0$ was applied. Finally, the role of jets, mini-jets and the effect of quarks originating from the incoming parton distribution functions needs to be quantified. The conclusions of this study are strongly suggestive, but are certainly not fully rigorous. That will require a full Bayesian analysis of both LHC and RHIC data. Such an effort will need to account for the theoretical uncertainties or model shortcomings mentioned above. A Bayesian analysis would be greatly enhanced by higher-statistics data and by having experimental results binned by both relative rapidity and relative azimuthal angle. There is good reason to believe such an analysis could be completed over the next few years.

Thus, the outlook for further analysis is bright as it appears that both fundamental questions above can be isolated in rather robust ways using balance functions. The work needed to improve the analyses and strengthen conclusions seems fairly clear. However, in addition to the need for additional work, there is the need for additional eyes on the problem. This field would benefit from having a broader slice of the field question the methods and inferences, both from the experimental and theoretical communities. Given the fundamental nature of the questions being addressed, this seems most warranted.

\begin{acknowledgments}
This work was supported by the Department of Energy Office of Science through grant number DE-FG02-03ER41259 and through grant number DE-FG02-87ER40328.  C.P.is supported by the US-DOE Nuclear Science Grant No. DE-SC0020633.
\end{acknowledgments}

\end{document}